\documentclass[aip,nofootinbib ]{revtex4}
 \pdfoutput=1
\usepackage{sidecap}
\usepackage[latin9]{inputenc}
\usepackage{graphicx}
\usepackage{amsmath}
\usepackage{amssymb}
\usepackage{amsfonts}
\usepackage{color}
\usepackage[super]{nth}
\usepackage{subfigure}
\usepackage[section]{placeins}

\makeatletter
\pdfpageheight\paperheight
\pdfpagewidth\paperwidth
\pdfoutput=1

\newcommand{\half}{\frac{1}{2}}
\newcommand{\quarter}{\frac{1}{4}}
\newcommand{\chris}[3]{\Gamma^{#1}_{\phantom{#1} #2 #3}}
\newcommand{\chrisg}[4]{#4^{#1}_{\phantom{#1} #2 #3}}

\newcommand{\vev}[1]{\left\langle #1 \right\rangle}

\newcommand{\scri}{\mathcal{I}}

\newcommand{\brak}[1]{\left(#1\right)}

\newcommand{\ssubfigure}[4]{\subfigure[#3]{
\includegraphics[width=#2\textwidth]{#1}
\label{#4}
}}

\begin{document}

 \begin{abstract}

Near horizon geometries have been widely studied, and have found many applications. Certain static, near horizon geometries are now understood to be bulk duals to CFTs with static scale-invariant sources under the AdS/CFT correspondence. However, static near-horizon geometries aren't just scale-invariant, they have extra `enhanced' symmetry. This means that they can only be the bulk duals for a special class of static, scale-invariant sources that share this enhanced symmetry. The purpose of this paper is to consider bulk duals for more generic static, scale-invariant sources, without this extra symmetry. These solutions are quite different to near-horizon geometries. In place of the extremal horizon they have a null singularity. We find specific examples of such bulk geometries numerically for the cases of pure gravity, and for an abelian gauge field.
\end{abstract}
\title{Bulk Duals for Generic Static, Scale-Invariant Holographic CFT States}
\author{Andrew Hickling}
\affiliation{Theoretical Physics Group, Blackett Laboratory, Imperial College, London SW7 2AZ, U.K. }
\email{a.hickling12@imperial.ac.uk}
\maketitle
\section{Introduction}
\label{sec:intro}
The AdS/CFT correspondence \cite{Maldacena1997,Witten:1998qj,Aharony1999} provides the opportunity to explore a particular class of strongly coupled field theories both in flat space, and on curved backgrounds. 
The dynamics of the CFTs that can be described by this correspondence is found, in the limit of large coupling and large numbers of degrees of freedom, by studying classical supergravity for spaces that are asymptotically locally $AdS_{n+1} \times X$, where $X$ is some compact internal space. The conformal field theory lives on the $n$ dimensional conformal boundary of this space. For given fixed choices of internal space $X$, we can always truncate to the Einstein's equations, $R_{\mu \nu}^{(n)} = \Lambda g_{\mu \nu}^{(n)}$, on an $n+1$ dimensional asymptotically AdS subspace. We will take a bottom-up view, and also consider the addition of an abelian gauge-field, without worrying about whether it is a consistent truncation of supergravity, as is often done in AdS/CMT \cite{Hartnol:2009}. When we do this we use the Einstein-Maxwell equations
\begin{equation}
\label{eq:einstsector}
\begin{split}
R_{\mu \nu}^{(n)} &= \Lambda g_{\mu \nu}^{(n)}+2 \left(F_{\mu \alpha}F_{\nu}^{\phantom{\nu} \alpha}-\quarter g_{\mu \nu} \left(F_{\alpha \beta}F^{\alpha \beta}\right)\right) \\
\nabla_\mu F^{\mu \nu} &= 0.
\end{split}
\end{equation}

Restricting  ourselves to these equations corresponds to examining a sector of dynamics in the large $N$, strong coupling limit, where the only vevs are the stress tensor, and a $U(1)$ charge current which is sourced by the boundary value of the gauge field. This correspondence has been used in many contexts to probe strongly coupled field theories, both on flat and curved backgrounds. For a review of the use of AdS/CFT to study strongly coupled CFTs on curved backgrounds, see \cite{Marolf:2013ioa}.

A particular subclass of the solutions to (\ref{eq:einstsector}) are the near-horizon geometries of  extremal horizons. These are reviewed and classified in \cite{Kunduri2013}. An important property of these solutions is their scaling symmetry, so, if they are asymptotically AdS, they can describe scale invariant states in a holographic CFT.  Let us restrict to the static case, so that from the point of view of the CFT these describe static, scale-invariant states. As proven in  \cite{Kunduri:2007vf}, static near horizon geometries have an enhanced SO(2,1) isometry group\footnote{Actually, what they show is that the near horizon geometry is a warped product of a maximally symmetric 2-d space, so the isometry group is the isometry group of either Minkowski, AdS, or dS. For the cases we're interested in here, it's AdS, with isometry group SO(2,1).}. This means they don't describe the most generic CFT states in this symmetry class.

For the remainder of this paper we specialize to 3+1 dimensions, but we expect that our results should generalize. In this case, all the static near-horizon geometries that solve (\ref{eq:einstsector}) are known analytically \cite{Kunduri2013}. The asymptotically AdS solutions have boundary metrics and electric potentials of the form\footnote{Note that the metrics in \eqref{eq:assymcone} are only determined up to a conformal factor. We can write $g$ in the conformal frame $$g = -dt^2 + dr^2 + \alpha^2 r^2 d\phi^2,$$ where we can interpret them as cones, or as flat space when $\alpha=1$. The electric source is invariant under this change of frame.}
\begin{equation}
 \label{eq:assymcone}
\begin{split}
 g & =-dt^2 + dr^2 + \alpha^2 d\phi^2 \frac{-dt^2 + dr^2}{r^2} + \alpha^2 d \phi^2 \\
 A & = e \frac{dt}{r}.
\end{split}
\end{equation}
In $3+1$ dimensions, we see that the isometry group is in fact $SO(2,1)\times U(1)$, with an additional $U(1)$ rotational symmetry in the $\phi$ coordinate. 
 
We might, instead, want to consider the CFT with sources that explicitly break this  $SO(2,1)\times U(1)$, either by introducing a boundary metric of the form
\begin{equation}
\label{eq:assymgen}
g \sim \frac{-dt^2 + dr^2}{r^2} + 2 \chi(\phi)\frac{dr}{r} d\phi + \alpha^2 d \phi^2,
\end{equation}
or by introducing an electric source with angular dependence
\begin{equation}
\label{eq:assymvgen}
A = V(\phi) \frac{dt}{r}.
\end{equation}
Introducing either of these sources breaks both the rotational symmetry, and part of the $SO(2,1)$, so that the only symmetry we are left with is the static, scale-invariant part. Thus the bulk dual clearly cannot be a near horizon geometry. The purpose of this paper is to consider these more generic bulks associated with static, scale-invariant states. As we will see, these solutions are quite different to near-horizon geometries, in particular the extremal horizon is replaced by a singular null surface.

Finding these bulk duals involves finding $3+1$ dimensional static scale-invariant solutions to  \eqref{eq:einstsector} with negative cosmological constant. Once you impose the symmetries, this becomes an effectively two-dimensional problem that can be solved numerically, using the Harmonic Einstein method\cite{Headrick:2009pv,Adam2011}. We are able to solve for a range of choices of $\chi(\phi)$ and $V(\phi)$ on the boundary. For the case of pure gravity (no gauge field), these solutions were conjectured to exist in \cite{Hickling2014}. They were also constructed to linear order as a perturbation to AdS. Here we demonstrate that the full, non-linearized solutions exist by constructing explicit examples.

The applicability of these results is not limited to scale-invariant boundary states. In \cite{Hickling2014} it was found that the near horizon geometries of extremal horizons could describe the large scale structure of holographic CFTs on geometries that are only asymptotically scale-invariant\footnote{The conformal freedom means that asymptotically scale-invariant metrics include asymptotically locally flat metrics.}. Specifically, it was found that if a static bulk contained an extremal horizon that met the boundary at infinity, then the near horizon geometry of this extremal horizon was determined by the geometry of null infinity on the boundary. In particular, static perturbations to the boundary geometry that fall off fast enough towards infinity would not affect the extremal horizon in the bulk, and so vacuum states would have the same large scale structure determined by this near-horizon geometry. These arguments were specifically for the case of pure gravity, however in \cite{Horowitz2014} solutions have been found numerically where an electric potential is introduced on the boundary which is not scale-invariant, but is asymptotically of the form (\ref{eq:assymcone}). In this case the solutions found again had the near-horizon geometry which was determined solely by this asymptotic form.

We would suggest that the appropriate way to describe the large scale structure of static CFT states on geometries that are asymptotically of the generalized form (\ref{eq:assymgen}), or where we have introduced electric potentials that are asymptotically of the form (\ref{eq:assymvgen}), is through the bulks we describe here. In particular, these geometries would be suitable for the infra-red of the bulk duals to states on boundaries of this form.

\section{Static Scale Invariant Geometries}
In Appendix \ref{sec:scaleinvapp} we argue that a general static, scale-invariant, $2+1$ dimensional boundary metric and electric potential can be written in the form
\begin{equation}
\label{eq:twistedboundary}
\begin{split}
g &= -\frac{dt^2}{r^2} + \frac{dr^2}{r^2} + \alpha^2 d\phi^2 + 2 \alpha \chi(\phi) d\phi \frac{dr}{r} \\
A &= V(\phi) \frac{dt}{r},
\end{split}
\end{equation}
where the scaling symmetry is given by
\begin{equation}
\label{eq:scalingsymmetry}
\begin{split}
t &\rightarrow \epsilon t \\
r &\rightarrow \epsilon r.
\end{split}
\end{equation}

Writing the metric in this form involves making a choice of coordinate frame. We can think of these static, scale invariant geometries as describing the large scale limit of boundary geometries with an asymptotic region of the form 
\begin{equation}
\label{eq:twistedboundaryrho}
g = - dt^2 + dr^2 + \alpha^2 r^2 d\phi^2- 2r \alpha \chi(\phi) d\phi dr,
\end{equation}
by Weyl scaling the metric by a factor of $r^2$. In this frame \eqref{eq:twistedboundaryrho}, the surface $r,t\to\infty$ is null infinity, $\scri^+$. If $\chi(\phi)=0$, it is an extremal horizon in \eqref{eq:twistedboundary} corresponding to the Poincare horizon of the $AdS_2$ factor. However, if $\chi(\phi)\neq0$, this surface is singular. Pragmatically, this can be seen because the metric in (\ref{eq:twistedboundaryrho}) cannot by transformed into Gaussian Null Coordinates, and this should be possible for any smooth hypersurface\cite{Moncrief1983}. This is the same singularity as we will see in the bulk of our solutions. Even if $\chi(\phi)=0$, turning on a non-constant $V(\phi)$ will generate the singularity in the bulk. The nature of this singularity, and the resulting singularity in the bulk solutions, will be explained more physically in Section \ref{sec:singularity}.

\subsection{Extremal Near Horizon Solutions}
\label{sec:extremalcase}
On the other hand, when we consider boundaries with $\chi(\phi)=0$, and constant $V(\phi) = e$, there are non-singular bulk solutions that preserve the Symmetry (\ref{eq:scalingsymmetry}). The boundary geometry in these cases is a cone with opening angle determined by $\alpha$. Alternatively, we can reinterpret this in the conformally compactified frame given in (\ref{eq:twistedboundary}). From this point of view it is $AdS_2 \times S^1$, and the parameter $\alpha$ now determines the relative sizes of the $AdS_2$ and $S_1$ factors.
These solutions are known analytically and they are the $3+1$ dimensional asymptotically AdS near horizon geometries\cite{Chrusciel2006}
\begin{equation}
\label{eq:4dSolution}
\begin{split}
g_{NH} &= \frac{d\psi^2}{f(\psi )}+\alpha^2 f(\psi) d \phi^2+\frac{\psi ^2 \left(dr^2-dt^2\right)}{r^2}\\
A &= e \frac{dt}{r}
\end{split}
\end{equation}
where $f(\psi) = \psi^2-1 -\frac{c}{\psi}- \frac{e^2}{\psi^2}$, and where  $\alpha$, $e$ and $c$ are constants. We have taken $\Lambda=-3$. These solutions run from $\psi_0 \leq \psi < \infty$, where $\psi_0$ is the largest zero of $f(\psi)$, and the conformal boundary is at $\psi \rightarrow \infty$, with metric
\begin{equation} g = \frac{-dt^2+dr^2}{r^2} + \alpha^2 d \phi^2.\end{equation} 
The constant $\alpha$ is fixed in  terms of $e$ and $c$ by requiring there to be no conical deficit in the bulk, and determines the conical deficit in the boundary metric.

The special case $\psi_0=\alpha=1$ with $e=0$ corresponds to pure $AdS_4$ in the bulk, and the conformal boundary is asymptotically Minkowski with no conical deficit and no charge density. For the solutions we find here, we are either going to deform the boundary while keeping $A=0$, or keep the boundary flat with no conical deficit while introducing an electric potential. We therefore elaborate on the extremal solutions in these two cases.

For the case $e=0$, we can write $c=\psi_0(\psi_0^2 -1)$. $\psi_0$ is the largest zero of $f(\psi)$ so long as $1/\sqrt{3}<\psi_0$. The parameter $\alpha$ which keeps the bulk conical-singularity free can then be found to be $\alpha = \frac{2 \psi_0}{1-3 \psi_0^2}$, so we have a one parameter family of solutions corresponding to the full range of possible values of $\alpha^2$.

If we instead impose $\alpha=1$, so the boundary is Minkowski, and allow a non-zero electric potential, we find another one parameter family of near horizon geometries. These solutions are discussed in \cite{Horowitz2014}, where they are interpreted as point charge defects in the CFT. We find $e$ in terms of the largest zero $\psi_0$ to be
\begin{equation}
\label{boundarygaugeformula}
e^2 = \psi_0^2 (1+3 \psi_0)(1-\psi_0).
\end{equation}
In this case $\psi_0$ is the greatest positive zero of $f$ so long as $0.226 \lessapprox \psi_0<1$. The parameter $\psi_0$ in this range labels a family of bulk solutions, and the dual CFT states are sourced by an electric potential $e$. As pointed out in \cite{Horowitz2014}, the map, $e \to \psi_0$, from source to state is neither left-complete nor one-to-one. The range of $e$ for which there is a corresponding state is bounded by $e^2 < \frac{1}{288}\left(69+11\sqrt{33}\right)\equiv e^2_m$, and, for $e^2 \gtrapprox 0.066 \equiv e^2_c$, there are two branches of states labelled by two different values of $\psi_0$. The two branches are, 
\begin{itemize}
\item \emph{Branch 1}:  Runs from $0 < e^2 < e^2_m$. The $e^2=0$ case is pure AdS.
\item \emph{Branch 2}: Runs from $e_c^2 < e^2 < e^2_m$. It meets Branch 1 at $e^2 = e^2_m$. 
\end{itemize}

When we introduce a more general electric potential, we will again find two states for certain choices of sources. 

\section{Numerical Setup}
\subsection{Anzatz and Coordinate Fixing}
The task is to find 4 dimensional solutions of the Einstein-Maxwell equations, with conformal boundary given by \eqref{eq:twistedboundary}, which are static and preserve the symmetry in \eqref{eq:scalingsymmetry}. These solutions generalise the near horizon geometries of \eqref{eq:4dSolution}. We will take units where $\Lambda = -3$.

The symmetries reduce this to an effectively two dimensional problem, with dependence on only an angular coordinate which we call $\phi$ and a bulk coordinate which we call $X$, a compactified version of the $\psi$ coordinate from (\ref{eq:4dSolution}). The dependence of the metric on the other two coordinates, $r$ and $t$, are then fixed by the symmetries. We compactify the bulk coordinate, so that $X$ lies in the range $0 \leq X < 1$. $X$ and $\phi$ taken together describe a unit disk, which we call $\mathcal{M}$, with the boundary of the disk being the conformal boundary. 



For the bulk anzatz, we write down the most general metric and gauge field compatible with the symmetries
\begin{equation}
\label{eq:abstractanzatz}
\begin{split}
g &= \frac{1}{(1-X^2)^2}\left(S_1(X,\phi)\frac{dt^2}{r^2} + S_2(X,\phi)\frac{dr^2}{r^2}+2\frac{dr}{r}\omega(X,\phi) + {g_2}_{ij}(X,\phi)dy^i dy^j\right)\\
A &= S_3(X,\phi) \frac{dt}{r},
\end{split}
\end{equation}
where $y=(X,\phi)$. From the perspective of $\mathcal{M}$, $S_1$, $S_2$, and $S_3$ are scalars, $\omega$ is a one-form, and $g_2$ is a symmetric rank 2 tensor. Note that as we have explicitly factored out the divergent conformal factor, we can choose coordinates such that these are all smooth. 

We need to impose boundary conditions, and these need to fix the conformal boundary at $X=1$ to be of the conformal class of (\ref{eq:twistedboundary}), and fix the electric potential to this value as well. To achieve this, the boundary conditions will satisfy
\begin{equation}
\label{eq:boundaryconditions}
\begin{split}
S_1(1,\phi) &= S_2(1,\phi) = 4\gamma \\
\omega(1,\phi) &=  4\gamma (\chi(\phi)d\phi) \\
g_2(1,\phi) &=  4 ( \gamma \alpha^2 d\phi^2 + dX^2)\\
S_3(1,\phi) &= V(\phi),
\end{split}
\end{equation}
where we will choose the constant $\gamma$ as convenient.

After fixing this anzatz we still have the residual coordinate freedom
\begin{equation}
\begin{split}
\label{eq:coordinatefreedoms}
r &\rightarrow r \lambda(X,Y) \\
X & \rightarrow \tilde{X}(X,Y) \\
Y & \rightarrow \tilde{Y}(X,Y).
\end{split}
\end{equation}
 We fix this using the Harmonic Einstein method. This method, and numerical approaches to applying it are discussed in \cite{Wiseman2011,Adam2011}. The essence is to gauge fix by choosing a gauge condition of the form
\begin{equation}
\label{eq:harmonicgauge}
\xi(x)^\alpha \equiv g^{\mu \nu}\left(\chris{\alpha}{\mu}{\nu} - \chrisg{\alpha}{\mu}{\nu}{\bar{\Gamma}}\right) = 0,
\end{equation}
where $\bar{\Gamma}$ is some chosen reference connection. This gauge condition is not imposed directly, but instead we add a  $\nabla_{(\mu}\xi_{\nu)}$ term to the Einstein-Maxwell equations
\begin{equation}
\label{eq:einstdeturk}
\begin{split}
R_{\mu \nu}&= \Lambda g_{\mu \nu}+2 \left(F_{\mu \alpha}F_{\nu}^{\phantom{\nu} \alpha}-\quarter g_{\mu \nu} \left(F_{\alpha \beta}F^{\alpha \beta}\right)\right) + \nabla_{(\mu}\xi_{\nu)}\\
\nabla_\mu F^{\mu \nu} &= 0.
\end{split}
\end{equation}

This additional term makes the  Einstein equations elliptic for static spacetimes. In the absence of the gauge field, there is also a maximum principle that says so long as $\phi = g_{\alpha \beta}\xi^\alpha \xi^\beta$ vanishes on the boundary it vanishes everywhere. This is derived in \cite{Figueras:2011va} by considering the contracted Bianchi identity. If the metric satisfies \eqref{eq:einstdeturk} with $F_{\mu \nu} =0$, this can be manipulated into the condition 
\begin{equation}
\label{phirelation}
 \nabla^2 \phi + \xi^\mu \partial_\mu \phi = - 2 \Lambda \phi + 2 (\nabla^\mu \xi^\nu)(\nabla_\mu \xi_\nu)
\end{equation}

In the static case, both terms on the right hand side are positive everywhere. This leads to the condition
\begin{equation}
 \nabla^2 \phi + \xi^\mu \partial_\mu \phi \ge 0,
\end{equation}
which, it can be shown, implies that $\phi$ takes its maximum value on the boundary. Since, in addition, $\phi \ge 0$, this implies that if it vanishes on the boundary, it vanishes everywhere. From this it follows that if we solve \eqref{eq:einstdeturk} without a gauge field, with boundary conditions that ensure $\phi=0$ on the boundary, we should end up with solutions to Einstein's equations.


If $F_{\mu \nu} \neq 0$, we do not have such a maximum principle, and we will have to check that our gauge condition is indeed satisfied by the solutions we find to be sure that we are actually solving Einstein's equations. Na\"{i}vely, you'd have to check for the vanishing of $\xi$, or equivalently $\phi$, everywhere. The issue with this is however small it is in the numerical solutions, you might worry that it doesn't actually vanish completely. Fortunately, in some cases we can do something a little better. The addition of a stress tensor to Einstein's equations, such as that due to $F_{\mu \nu}$ in \eqref{eq:einstdeturk}, modifies the relation \eqref{phirelation} to
\begin{equation}
\label{eq:generalphirelation}
\nabla^2 \phi + \xi^\mu \partial_\mu \phi = - 2 \Lambda \phi + 2 (\nabla^\mu \xi^\nu)(\nabla_\mu \xi_\nu) - 2 \tilde{T}_{\mu \nu} \xi^\mu \xi^\nu,
\end{equation}
where $\tilde{T}_{\mu \nu} = T_{\mu \nu} + \frac{T}{2-d}g_{\mu \nu}$. Note that in the static case, we can choose $\bar{\Gamma}$ such that $\xi$ is spacelike, so that  if the spacial part of $-\Lambda g_{ij}- \tilde{T}_{ij}$ is positive definite, the right hand side of this equation is again positive definite. We can't in general assume that this will be true, but we can test whether this matrix is positive definite for the solution that we get, and, if it is, we know for sure that we must have $\phi=0$. In principle, this can be a better test than directly testing $\phi=0$, because, so long as the eigenvalues of the matrix don't get too small, we can demonstrate that they are finite and positive without having to get into detailed arguments about convergence.  

\begin{figure}
\includegraphics[width=0.6\textwidth]{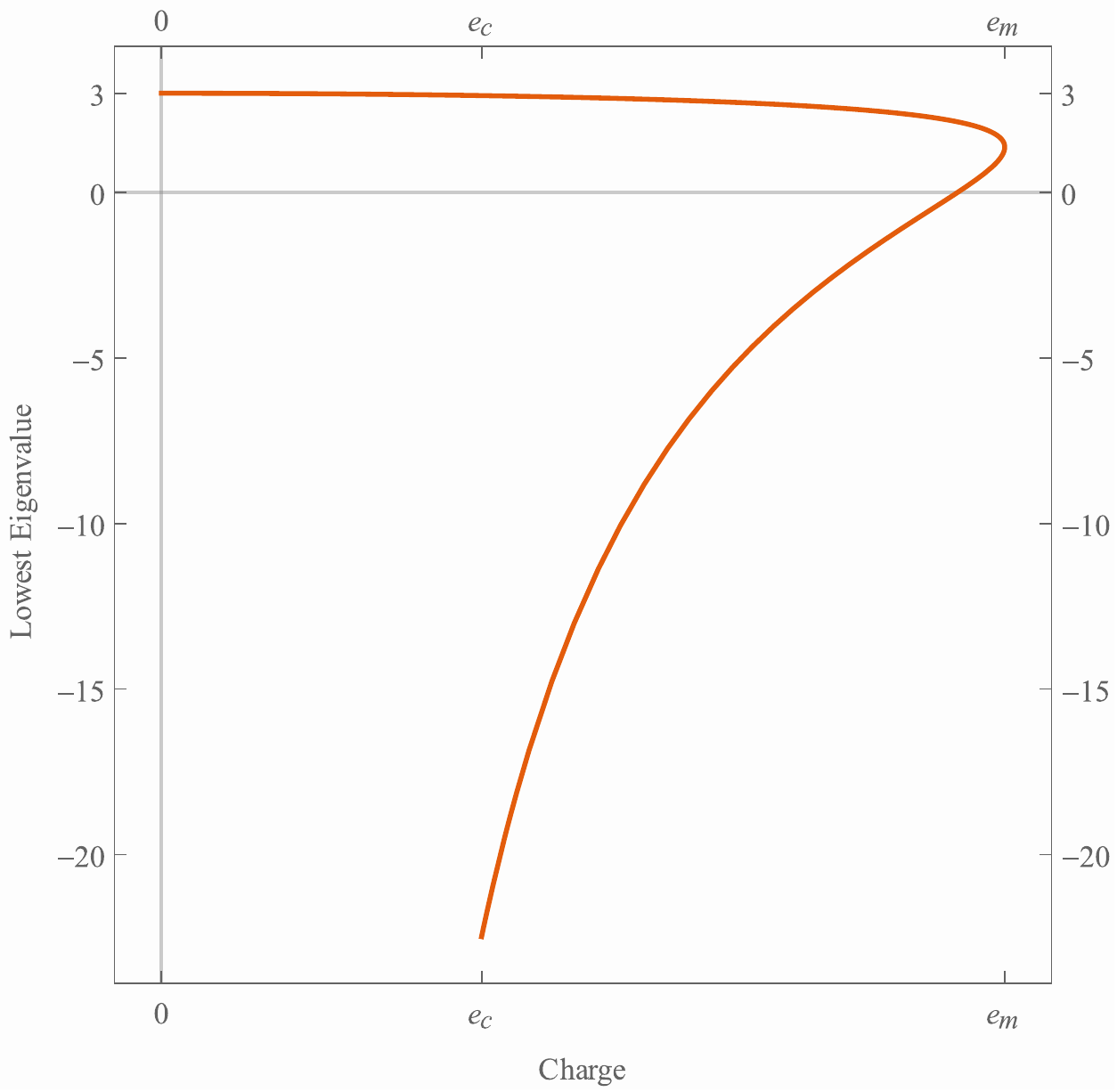}
\caption{Minimum value of the minimum eigenvalue of the spatial matrix $-\Lambda \delta^{i}_{j} + T^{i}_{j}$ for the near-horizon geometries as a function of charge. For the entirety of \emph{Branch 1}, as defined in \ref{sec:extremalcase}, this eigenvalue is positive, meaning that the matrix is positive definite.}
\label{fig:mineigenvalue}
\end{figure}

In Figure \ref{fig:mineigenvalue} we examine the positive definiteness for this matrix for the near horizon geometries. We see that throughout \emph{Branch 1}, as defined in Section \ref{sec:extremalcase}, this matrix is positive definite, whereas it becomes negative in \emph{Branch 2}.  For the more general numerical solutions, this picture isn't changed too much, and there is a region of solutions where this matrix is positive definite, and a region where it isn't. For the cases where this is positive definite, this demonstrates that we have actually solved Einstein's equations, so long as $\phi$ vanishes on the boundary, and, for the other cases, we have to check $\phi$ directly.

 We will derive our reference connection $\bar{\Gamma}$ from a reference metric $\bar{g}$ which we will take to have the same general form as the metric \eqref{eq:abstractanzatz}. So long as we fix the conformal boundaries of $g$ and $\bar{g}$ to be the same, it follows, by examining the expansion of the condition \eqref{eq:harmonicgauge} near $X=1$, that we will have  $\phi=0$ on the boundary. We will therefore pick a reference metric which also satisfies \eqref{eq:boundaryconditions}. The precise choice will depend on the values of $\alpha$, $\chi(\phi)$ and $V(\phi)$.

Note that there is in general no guarantee that, for a given choice of $\bar{g}$, the coordinate choice implied by the condition \eqref{eq:harmonicgauge} is smooth. Indeed, as we show in Appendix \ref{sec:logarithm}, in the case with a gauge field we can show that at the boundary our choice of coordinates is only smooth up to fourth derivative. That said, evidence from the analysis of the convergence of our solutions, presented in Appendix \ref{sec:convergence}, suggests that our solutions in these coordinates are in fact $C^4$.

\subsection{Discretization}
We have reduced the problem to a set of  two dimensional pde's, depending on X and $\phi$, which should be thought of as polar coordinates on a unit disk with boundary conditions given by \eqref{eq:boundaryconditions}. Now we need to construct a lattice on which to discretize the problem.  For simplicity, we would like to take a rectangular lattice, and we would like the conformal boundary, $X=1$, to be one edge of this lattice. This leads us to using the polar $X,\phi$ coordinates themselves as the lattice coordinates. 

These coordinates, of course, break down at the origin $X=0$. In order to avoid this issue, we limit our attention to solutions that are invariant under parity
\begin{equation}
\label{eq:parity}
\phi \rightarrow \phi + \pi 
\end{equation}
To take advantage of this, we cover the disk twice by extending the $X$ coordinate $-1<X<1$. The extension is defined such that the map from polar to Cartesian coordinates is generalised to hold for negative $X$
\begin{equation}
\label{eq:polar}
\begin{split}
x &= X \cos \phi \\
y &= X \sin \phi.
\end{split}
\end{equation}

These coordinates map out the disk twice, and consistency then requires that when we extend our metric and gauge field to this double covering, they are invariant under
\begin{equation}
\label{eq:doublepolarconsistency}
\begin{split}
X&\rightarrow-X \\
\phi &\rightarrow \phi+\pi,
\end{split}
\end{equation}
as this sends $x\to x$ and $y \to y$. Imposing this consistency relation would be messy, but, given our assumed parity symmetry \eqref{eq:parity}, the solutions are in fact symmetric under \emph{each} of $X \to -X$ and $\phi \to \phi+\pi$ separately.

Imposing each of these conditions separately is straight forward. We construct a lattice where $X$ runs from $-1$ to $1$, with an even number of points so that we don't need to place a point at $X=0$, and $\phi$ runs from $0$ to $\pi$. We construct a $\nth{6}$ order finite difference matrix in the $X$ direction, and then identify the points under $X\rightarrow -X$ so that we have resulting differentiation matrices for the points $X>0$ which builds in the evenness without having a point at the origin which would require special treatment. In the $\phi$ direction we use Fourier differencing to build in the $\pi$ periodicity.

In order for this to make sense, we need to choose an anzatz for the metric and gauge field in which the functions that define the components of both inherit the $X \to -X$ and $\phi  \to \phi+\pi$ symmetries. Such an anzatz is given by
\begin{equation}
\label{eq:specificanzatz}
\begin{split}
S_1(X,\phi) &= (1+X^2)(1+A(X,\phi))\\
S_2(X,\phi) &= (1+X^2)(1+B(X,\phi))\\
\omega(X,\phi) &= F(X,\phi) d(X^2) + X^2 H(X,\phi) d\phi\\
g(X,\phi) &= 4(1+L(X,\phi))(dx^2+dy^2) + 2M(X,\phi)dx dy + S(X,\phi)(dx^2-dy^2)\\
S_3(X,\phi) &= V(X,\phi),
\end{split}
\end{equation}
with $x = X \cos \phi$ and $y = X \sin \phi$.  In addition to the symmetry, this anzatz has two nice properties. Firstly, if all the functions vanish we are left with pure AdS. Secondly, a simple sufficient condition for these objects to be smooth on $\mathcal{M}$ is that the functions themselves should be smooth functions on $\mathcal{M}$. Note that, although we have avoided the numerical issues of having a point at the origin itself, $X$ and $\phi$, being polar coordinates, are still not good coordinates at the origin. In order to test the smoothness of the functions at this point, we will need to transform back to the Cartesian coordinates $x$ and $y$. This is elaborated on in Appendix \ref{sec:convergence}.

Having discretized the problem, we will find solutions using Newton itteration. An initial guess at a solution is made, and then we repeatedly improve our guess by solving linear systems. The linear systems are solved using UMFPACK\cite{suitesparse}.

\subsection{Choice of Boundaries to consider}

Even after imposing \eqref{eq:parity}, the space of possible boundary metrics and gauge fields of the form (\ref{eq:twistedboundary}) is infinite dimensional, so we need to focus on some particular subspace. We will separately consider the two cases where we deform the metric away from locally flat cones in the absence of a gauge field, and where we leave the boundary metric flat with no conical deficit, but introduce a non-isotropic electric potential. 

For the metric deformations, we choose the two dimensional subspace where
\begin{equation}
\chi(\phi) = \lambda \sin (2 \phi),
\end{equation}
and we shall look at varying values of $\alpha$ and $\lambda$. The one-dimensional sub-space, $\lambda=0$, is the set of boundary cones, for which, as described above, the bulks are the extremal solutions known analytically. Our reference metric needs to have this same deformed conformal boundary, so we take
\begin{equation}
\bar{g}_{\mu \nu}= \frac{1}{\left(1-X^2\right)^2}\left(h(X)\frac{\left(1+X^2\right)^2}{r^2} \left(-dt^2 + dr^2\right) + 4\left (dX^2 +X^2 d\phi^2\right) +  8\lambda X^2 k(X)\sin 2 \phi \frac{drd\phi}{\alpha r}  \right),
\end{equation}
with $h(X) = 1+\frac{1-\alpha^2}{\alpha^2}X^2$ and $k(X) = 1 - \left(1-X^2\right)^4$. In order to impose the boundary condition that $g_{\mu \nu}$ and $\bar{g}_{\mu \nu}$ should agree on the boundary, we impose the boundary condition \eqref{eq:boundaryconditions} with $V(\phi)=0$ and $\gamma =\alpha^{-2}$.

For the gauge field deformations, we will fix $\alpha = 1$ and $\chi(\phi)=0$, and we will again take a two dimensional subspace, this time
\begin{equation}
V(\phi) = a + b \cos(2 \phi).
\end{equation}
Again, the case $b=0$ are the boundaries for which we know the two branches of analytic extremal horizon solutions. In this case the boundary is unperturbed, so we could in principle use pure AdS as our reference metric. However, we choose instead to use the two branches of extremal horizon solutions as reference metrics, as this makes it easier to find solutions close to each of these branches. We therefore write the reference metric as

\begin{equation}
\bar{g}_{\mu \nu}= \frac{1}{\left(1-X^2\right)^2}\left( \psi_0^2\frac{\left(1+X^2\right)^2}{r^2} \left(-dt^2 + dr^2\right) + 4\left (\frac{dX^2}{P(X)} + P(X) \psi_0^2 X^2 d \phi^2\right)\right),
\end{equation}
with $P(X)=-\frac{\psi _0 \left(X^2-1\right)^3+X^2 \left(X^2-1\right)^2-2 \psi _0^2 X^2 \left(X^4-2 X^2+3\right)}{\psi _0^2 \left(X^2+1\right)^2}$. The parameter $\psi_0$ is the parameter that labels these near horizon solutions from Section \ref{sec:extremalcase}, and it depends on the value of $a$ through \eqref{boundarygaugeformula}. In order to impose the boundary condition that $g_{\mu \nu}$ and $\bar{g}_{\mu \nu}$ should agree on the boundary, we once again impose the boundary condition \eqref{eq:boundaryconditions}, this time with $\chi(\phi)=0$, $\alpha=1$ and $\gamma = \psi_0^2$.

Note that, in addition to being invariant under $\phi\to\phi+\pi$ as we required, the boundaries we've selected through our choice of $\chi(\phi)$ and $V(\phi)$ have a $\phi\to\pi-\phi$ symmetry. Since we don't break this symmetry in our choice of reference metric, and so in our choice of coordinates, it will be manifest in our solutions as well.

\section{Results}
\label{sec:results}

The results presented here have $25$ points in the $\phi$ direction and $145$ or $65$ in the radial direction for the boundary metric and electric potential deformations respectively\footnote{The memory requirements are higher with the addition of a gauge field.}. We already get good solutions at this resolution, but in Appendix \ref{sec:convergence} we demonstrate convergence by taking a subset of these solutions to higher resolution. For instance, in all regions of the solution space we explore, we are able in this way to push the error on the gauge condition, $\xi^{\mu}=0$, and on the Einstein's Equations down to $O(10^{-6})$. This is discussed in detail in the Appendix. These solutions are found for a range of the parameters $\lambda$ and $\alpha$ in the case of boundary metric deformations, and $a$ and $b$ in the case of boundary electric potential deformations. They were found using the facilities of the Imperial College High Performance Computing Service\cite{hpc}.

\subsection{Electric Potential Deformations}
The boundary charge can be written as\cite{Horowitz2014}
\begin{equation}
\label{eq:boundarycharge}
Q = \lim_{z\to 0} \frac{1}{4 \pi} \int_{\mathcal{S}(z)}*F,
\end{equation}
where $z$ is a function that goes to $0$ on the boundary, and $\mathcal{S}(z)$ is a surface of constant $z$ and $t$ that becomes a spacelike hypersurface in the boundary as $z\to 0$.
\begin{figure}
\centering
\def\svgwidth{0.5\textwidth}
\input{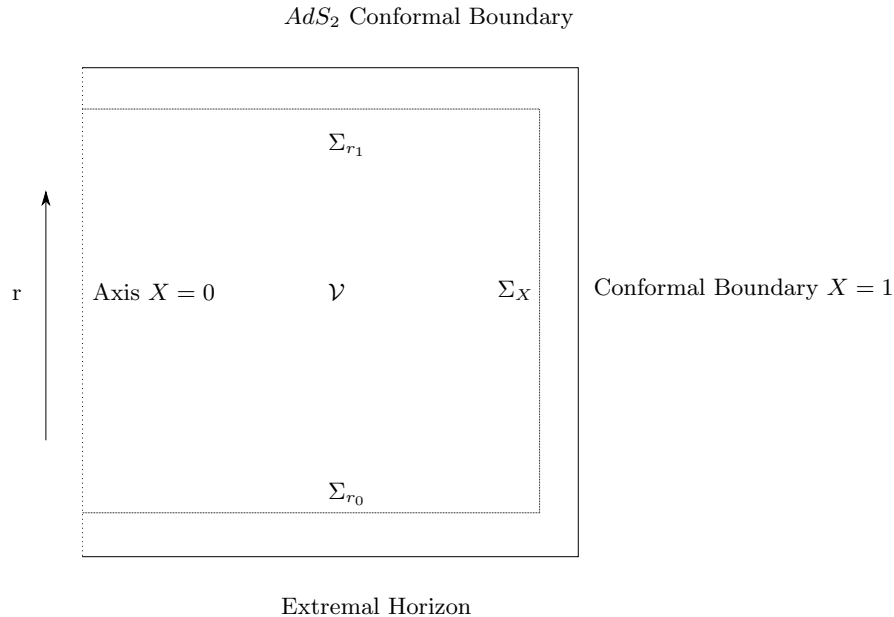}
\caption{A Constant Time slice of the Bulk}
\label{fig:chargeargument}
\end{figure}

We've chosen coordinates where the conformal boundary at $X=1$ is $AdS_2 \times S^1$, but we need to be careful when relating this back to coordinates where the conformal boundary is Minkowski. When we consider things more carefully, we find that the full conformal boundary is divided into two components, one at $X\to 1$ which is the main component of the conformal boundary, and another at $r \to \infty$,  the conformal boundary of the $AdS_2$ factor in these coordinates, which is the piece of the boundary corresponding to the origin in the Minkowski frame. So, in the transformation from the Minkowski frame to the $AdS_2 \times S^1$ frame, the origin has been blown up to the conformal boundary of the $AdS_2$.

A constant time, constant $\phi$ slice of our spacetimes is shown in Figure \ref{fig:chargeargument}. In this diagram, these two components of the conformal boundary are at the right hand side and the top. The full timeslice is the cylinder of revolution of this 2-d slice about the axis $X=0$. To calculate the charge as in \eqref{eq:boundarycharge}, we need to construct a sequence of surfaces $S(z)$ that become a timeslice of the conformal boundary as $z\rightarrow 0$.  Consider a cylinder, $\mathcal{V}$, within the spaceslice as in Figure \ref{fig:chargeargument}, where the top and bottom boundaries are at two values of $r$, $r_1$ and $r_0$, labelled $\Sigma_{r_1}$ and $\Sigma_{r_0}$ respectively, and the other surface of the cylinder, $\Sigma_X$, is at a fixed value of $X$. In the limit that we take this cylinder large, $\Sigma_X \cup \Sigma_{r_1}$ approaches both components of the conformal boundary. It is therefore a suitable choice to extract the boundary charge. We can write
\begin{equation}
\label{eq:boundarychargeb}
Q = \lim_{X \to 1 ,  r_1 \to \infty} \frac{1}{4 \pi} \left(\int_{\Sigma_{r_1}}*F+\int_{\Sigma_X}*F\right).
\end{equation}
These two terms represent the contribution of the two components of the conformal boundary to the total charge. From the point of view of the Minkowski frame,  the first term is a delta function charge density at the origin and the second term represents a charge distribution away from the origin. 

In fact, there is no overall contribution from this second term. We see this as follows. The Maxwell Equation can be written
\begin{equation}
d*F = 0.
\end{equation}
We can integrate this over the volume $\mathcal{V}$ using Stoke's theorem,
\newcommand{\sigint}[1]{\int_{\Sigma_{#1}}}
\begin{equation}
\int_\mathcal{V} d*F = \sigint{X}*F + \sigint{r_0} *F + \sigint{r_1}*F=0,
\end{equation}
where all the surfaces are oriented outward from $\mathcal{V}$. Our scaling symmetry in fact implies that
\begin{equation}
\sigint{r_1}*F = -\sigint{r_0}*F,
\end{equation}
so that,
\begin{equation}
\label{eq:xcharge0}
\sigint{X}*F = 0.
\end{equation}

\begin{figure}
\centering
\includegraphics[width=\textwidth]{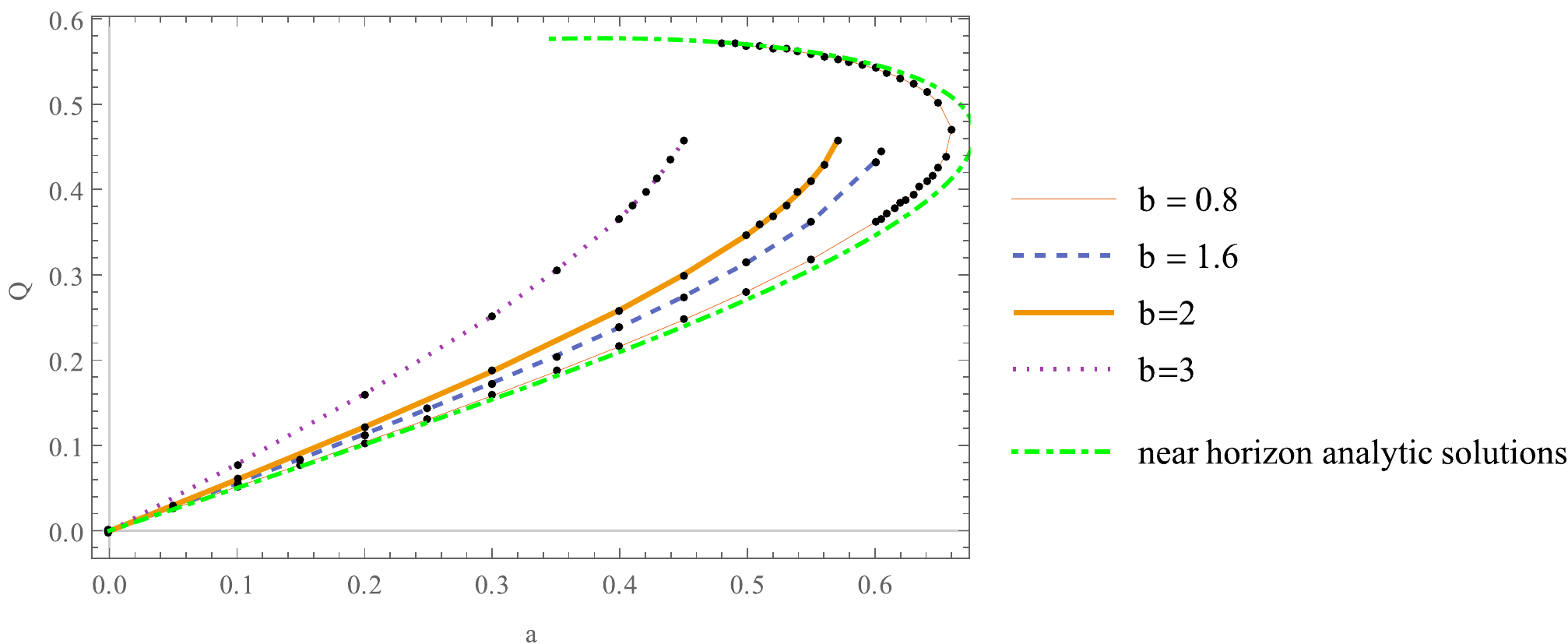}
\caption{Overall charge on solutions as a function of the parameters $a$ for a section of choices of $b$ in the electric potential source term $V(\phi) = a + b \cos 2\phi$ for which we find solutions. Note that at $a=0$ the total charge is always zero, which we would expect from symmetry.}
\label{fig:origincharge}
\end{figure}

This means that the only overall charge contribution can come from the origin, although there will be a non-zero charge density elsewhere on the boundary. In Figure \ref{fig:origincharge} we plot some examples of charges at the origin as a function of $a$ for various values of $b$ for the boundary source potential $V(\phi) = a+ b \cos 2 \phi$, together with the analytic curve for the near horizon solutions ($b=0$). These curves all go through the origin, which makes sense since, when $a=0$, there's nothing to pick out a preferred sign of charge. For each value of $a$ and $b$ we found these solutions by using a corresponding near-horizon solution with the matching value of $a$ as the reference metric and as our initial guess for the metric, and then proceeding with a simple Newton itteration. In the regions of values of $a$ where there are two branches of near-horizon solutions, we tried using each of these branches. In this way, we were able to find two solutions for a range of $a$ and $b$ pairs. For instance, we can see in the $b=0.8$ curve in Figure \ref{fig:origincharge}, a range of $a$ values for which, like in the near-horizon case, there are two branches of solutions. For the larger values of $b$ we weren't able to find two branches of solution in this way, but we have no reason to rule them out. Additionally, we weren't able to find solutions for values of $a$ greater than the maximum value on the near-horizon solutions.

In addition to the delta function charge, there is a non-zero charge density away from the origin, which in our case corresponds to a non-zero charge density on the $X=1$ conformal boundary. In the $AdS_2 \times S^1$ frame, because of the scaling symmetry, both the electric potential and charge density fall off as  $\frac{1}{r}$ , and have non-trivial dependence only on $\phi$. In the Minkowski frame this corresponds to a $1/r$ fall off for the potential, and a $1/r^2$ fall off of the charge density. From \eqref{eq:xcharge0} we expect the integral of this charge density to vanish, and in all of our numerical solutions this integral does not rise above $10^{-6}$. In this large scale region we also look at the induced energy density. This energy density is found via a Fefferman-Graham expansion\cite{Balasubramanian:1999re,Skenderis2000}, and, in the Minkowski frame, it falls off as $1/r^3$.

\begin{figure}
\centering
\ssubfigure{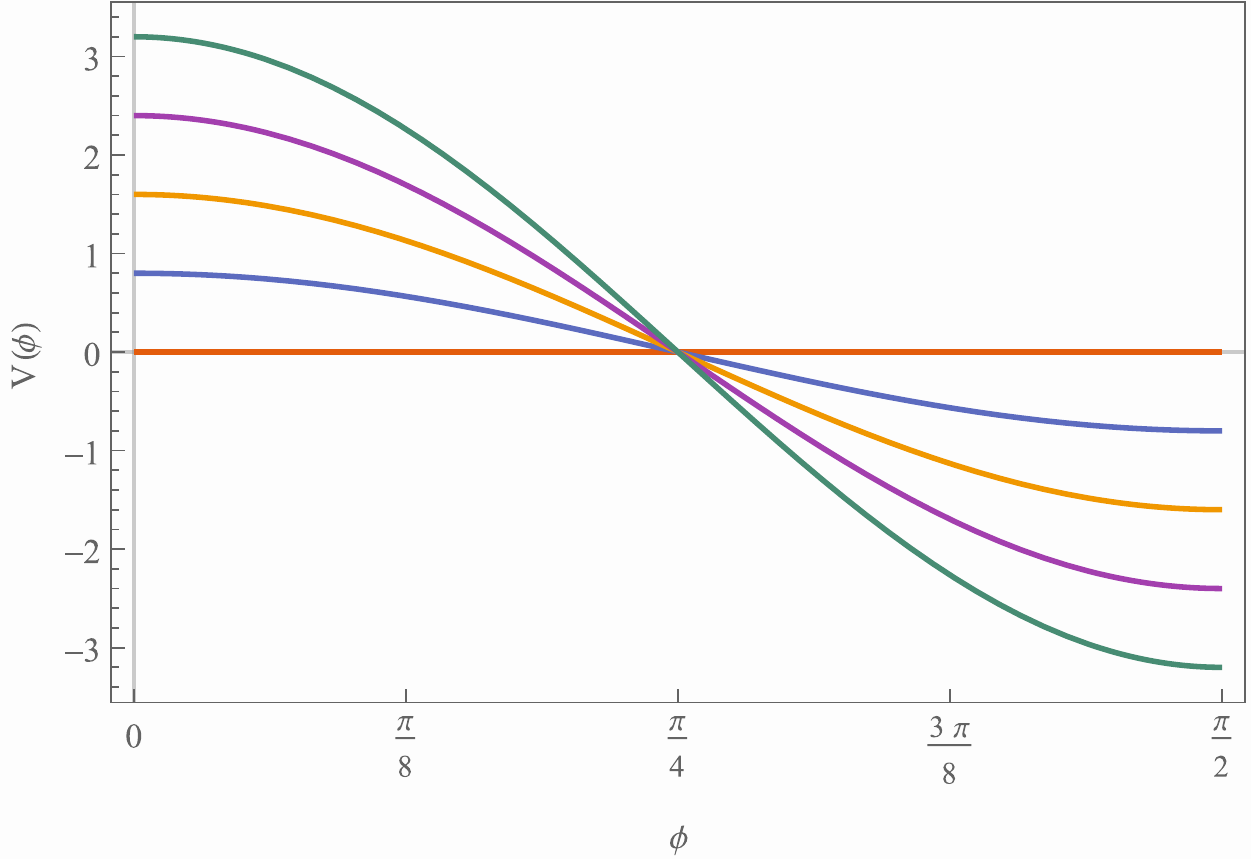}{0.4}{Form of the applied source $V(\phi)$ }{fig:epa0}
\ssubfigure{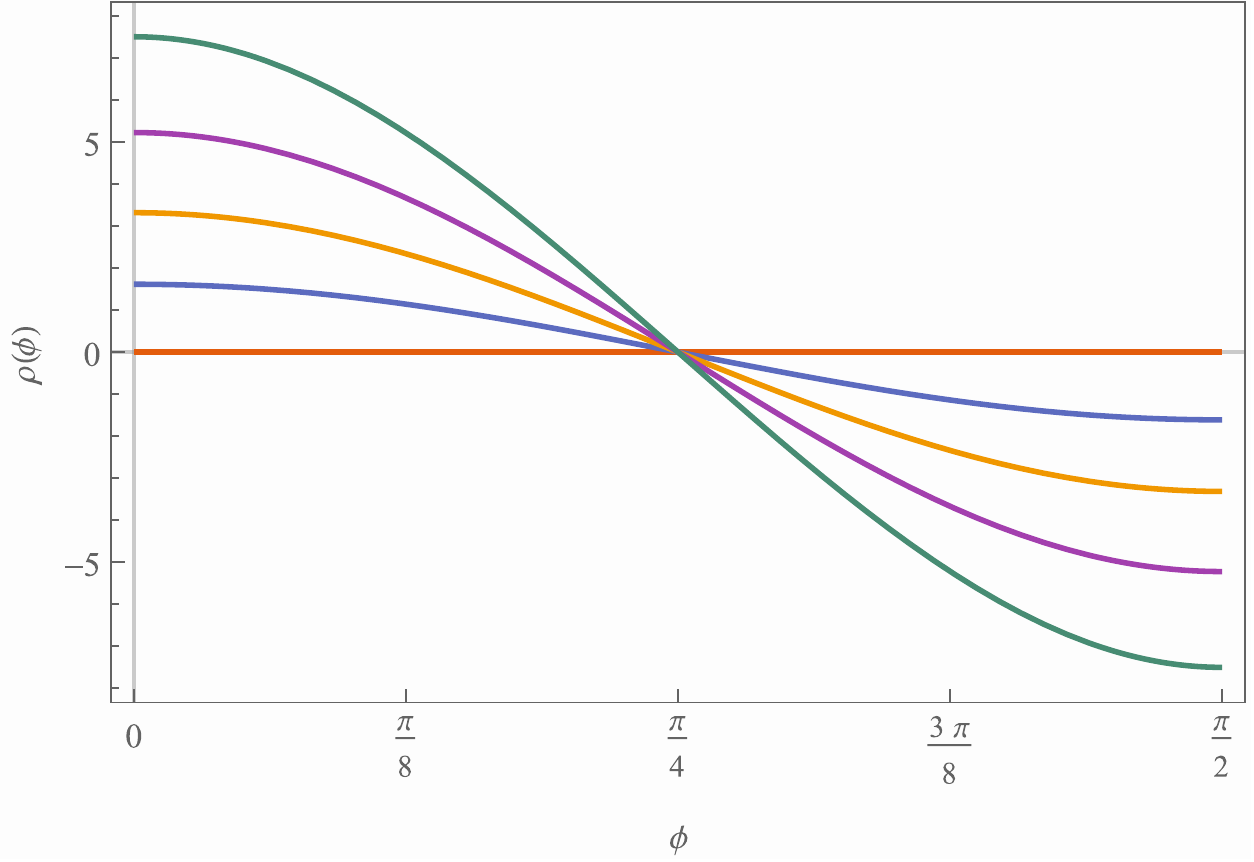}{0.4}{Resulting charge density $\rho(\phi)$ }{fig:cda0}
\ssubfigure{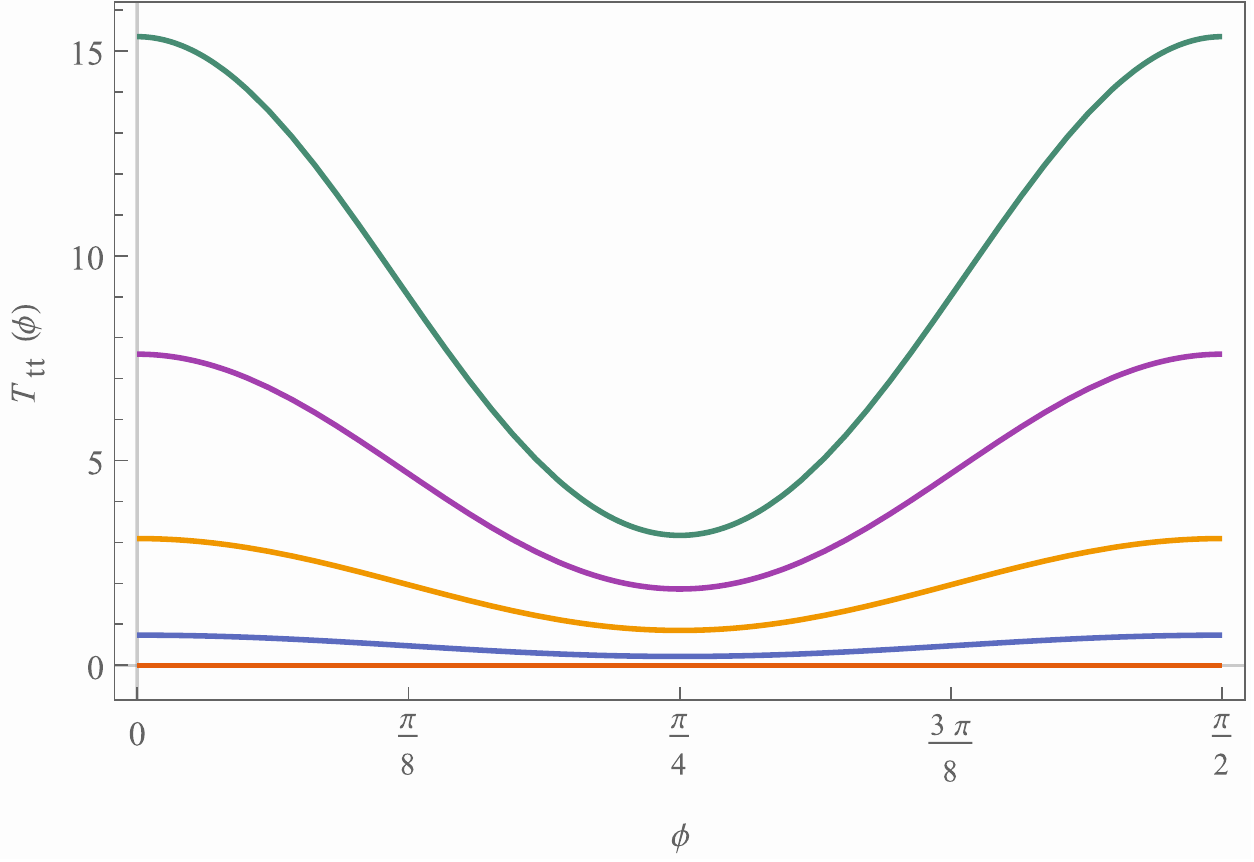}{0.4}{Resulting energy density $T_{tt}(\phi)$ }{fig:eda0}
\caption{Response to an electric potential source term $V(\phi) = \frac{b \cos 2\phi}{r}$, for $b=\left\{0,0.8,1.6,2.4,3.2\right\}$. The induced charge density integrates to zero, but the energy density is positive definite. Note that, quite intuitively, the energy density is peaked when the the magnitude of the charge density is largest. Also, there is a symmetry in the energy density between the regions of positive and negative charge, which makes sense because there is no overall charge at the origin. Viewed in a flat frame the potential goes as $1/r$, the charge density as $1/r^2$, and the energy density falls off as $1/r^3$, and we have evaluated all of them at $r=1$.}
\label{fig:largescaleresponsea0}
\end{figure}

\begin{figure}
\centering
\ssubfigure{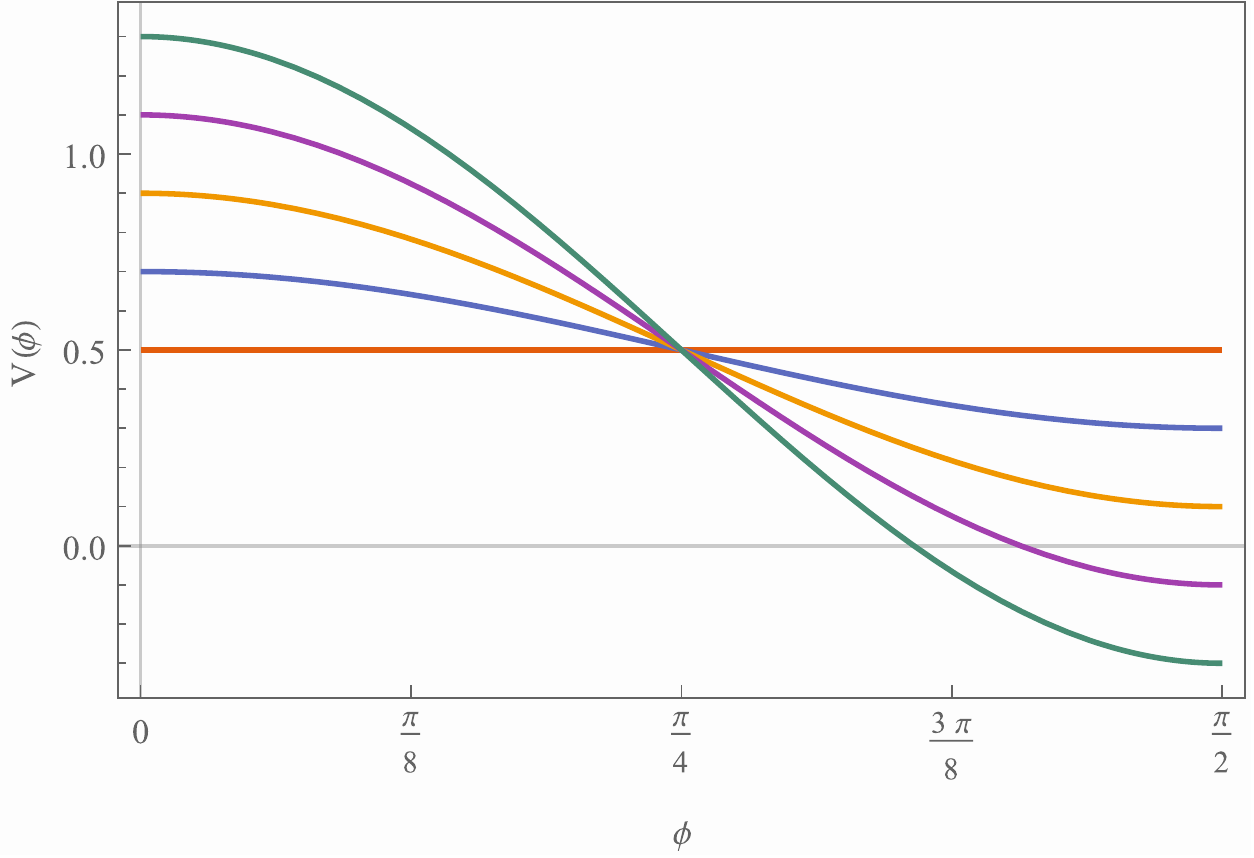}{0.4}{Form of the applied source $V(\phi)$ }{fig:epa5}

\centering
\ssubfigure{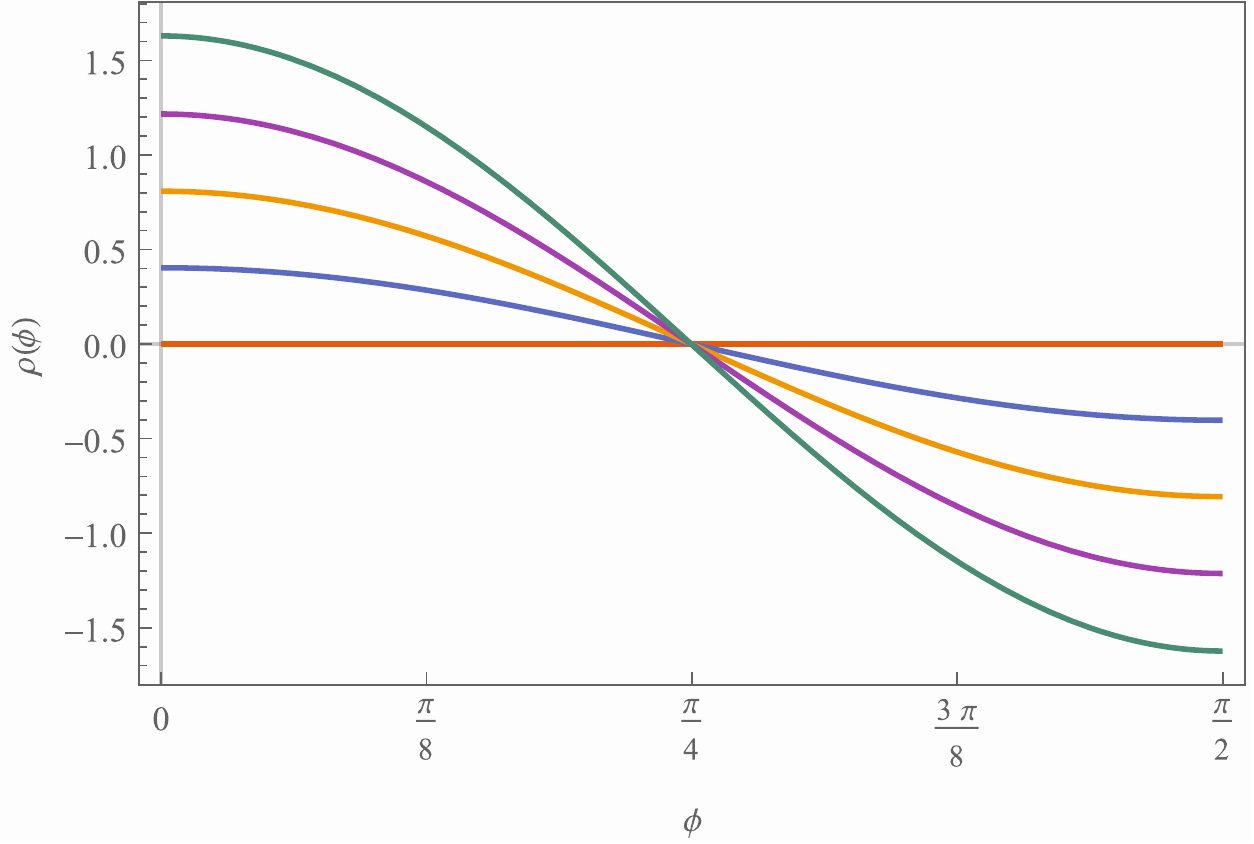}{0.4}{Resulting charge density $\rho(\phi)$ for branch 1}{fig:cda5}
\ssubfigure{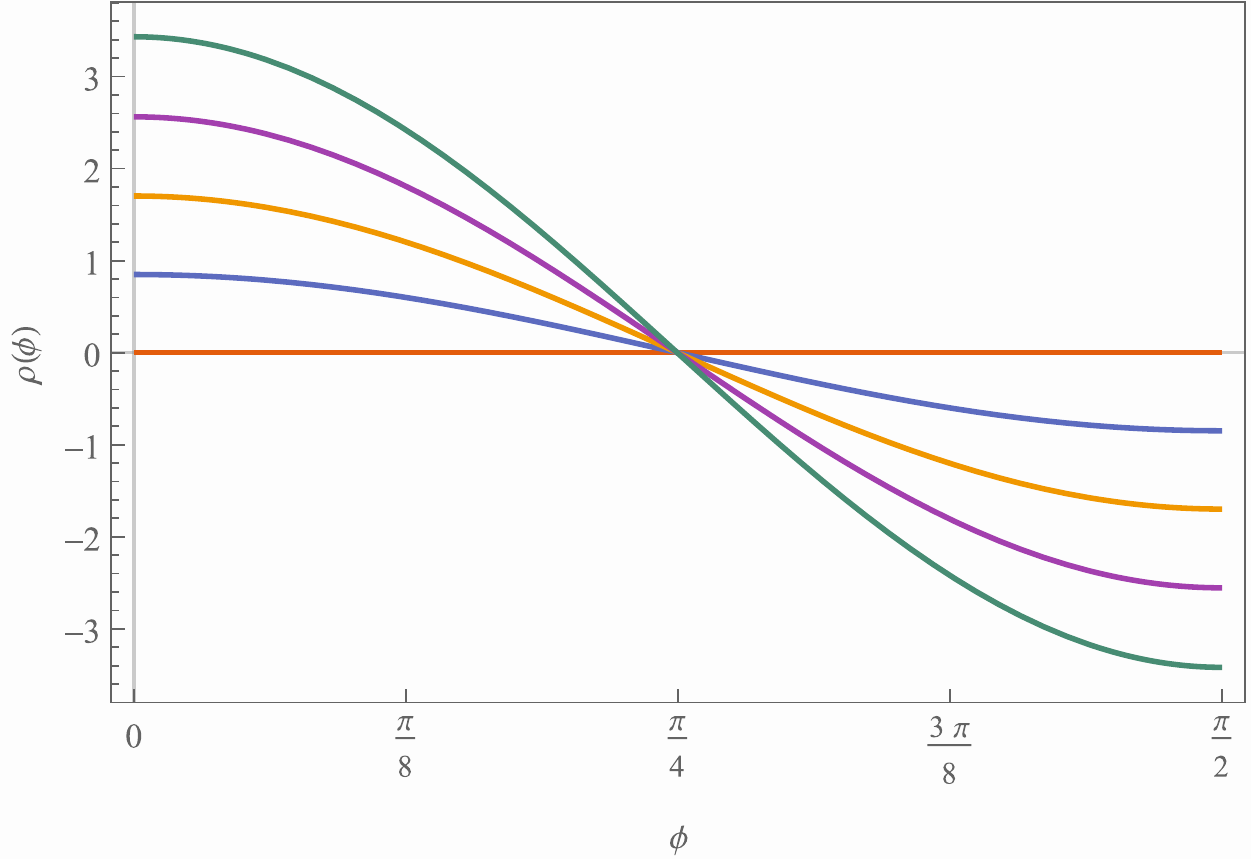}{0.4}{Resulting charge density $\rho(\phi)$ for branch 2}{fig:cda52}
\ssubfigure{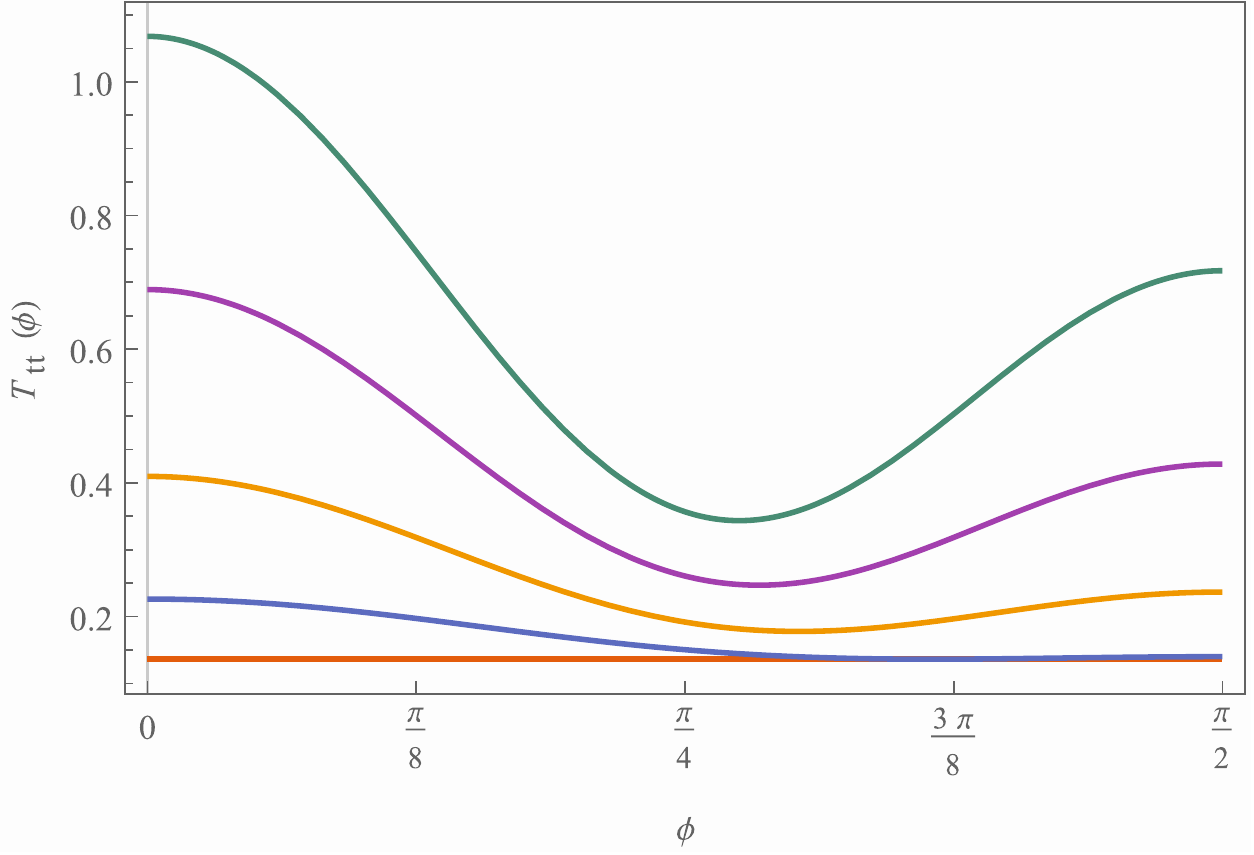}{0.4}{Resulting energy density $T_{tt}(\phi)$ for branch 1}{fig:eda5}
\ssubfigure{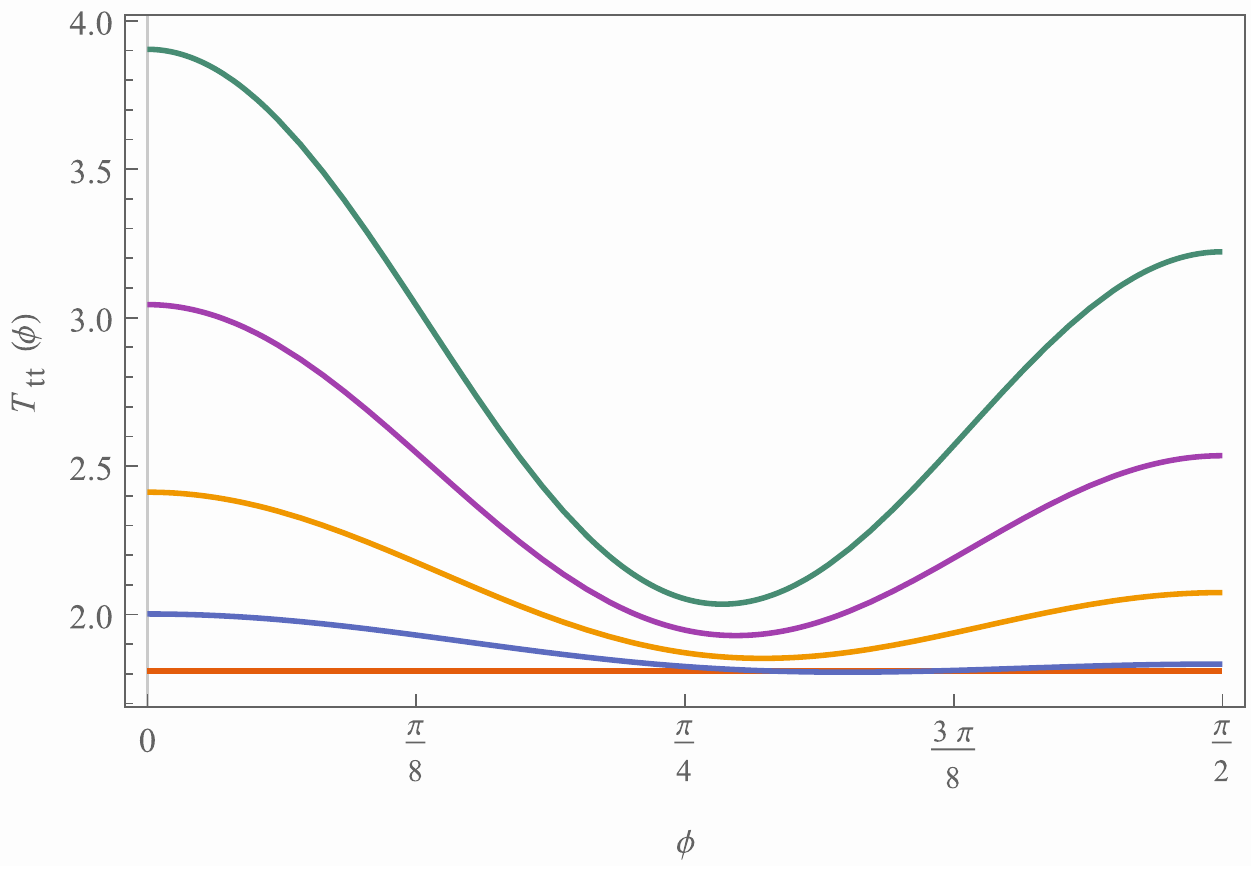}{0.4}{Resulting energy density $T_{tt}(\phi)$ for branch 2 }{fig:eda52}
\caption{Response to an electric potential source term $V(\phi) = \frac{0.5+b \cos 2\phi}{r}$ for $b=\left\{0,0.2,0.4,0.6,0.8\right\}$. In this case we found two branches of solutions. Note that even though the potential $V(\phi)$ is no longer centred about zero, unlike in Figure \ref{fig:largescaleresponsea0},the charge density still is, and it still integrates to zero. The main difference now is that we no longer have a symmetry in the energy density between regions of positive charge and regions of negative charge. This is because these solutions have an overall (positive) charge at the origin. Again, they have all been evaluated at $r=1$. }
\label{fig:largescaleresponsea5}
\end{figure}

We show plots from two sets of examples to illustrate how these field theory observables depend on the source potential. First, the $a=0$ case where $V(\phi) = b \cos 2 \phi$  for various values of $b$ . This is shown in Figure \ref{fig:largescaleresponsea0}. In this case there is only one branch of solutions, and there is a symmetry between the regions of positive and negative charge density because our boundary conditions don't pick out a sign of charge. The energy density is positive definite and, as you might expect intuitively, it's largest in regions of large density. If we consider instead $V(\phi) = 0.5+b \cos 2 \phi$, in Figure \ref{fig:largescaleresponsea5}, we find two branches of solutions. This time, because of the offset of $V(\phi)$ from zero, the symmetry between regions of positive and negative charge is broken. This can be seen most clearly in the energy density distribution which is larger in the region of positive charge density than in the region of negative density. Once again, the energy density is positive definite, and in fact seems to stay above the pair of constant values it takes in the corresponding $b=0$ extremal solutions. 

\subsection{Metric Deformations}

The other set of solutions we found were ones where we did not introduce a gauge field, and only deformed the boundary metric. In this case, the observable we focus on is the $X=1$ boundary stress-tensor, which we extract using a Fefferman-Graham expansion. We will focus on the energy density, which, as remarked above, goes as $1/r^3$. Whereas in the case of the electric potential deformations the energy densities were positive, once we deform the boundary  metric we get regions of negative energy density. This was already true for the near horizon solutions, described in Section \ref{sec:extremalcase}. All we do in that case is introduce a conical deficit on the boundary ($\chi(\phi)=0$). The resulting $T_{tt}(\phi)$ is a constant which is positive for $\alpha >0$ and negative for $\alpha < 0$.

\begin{figure}
\centering
\ssubfigure{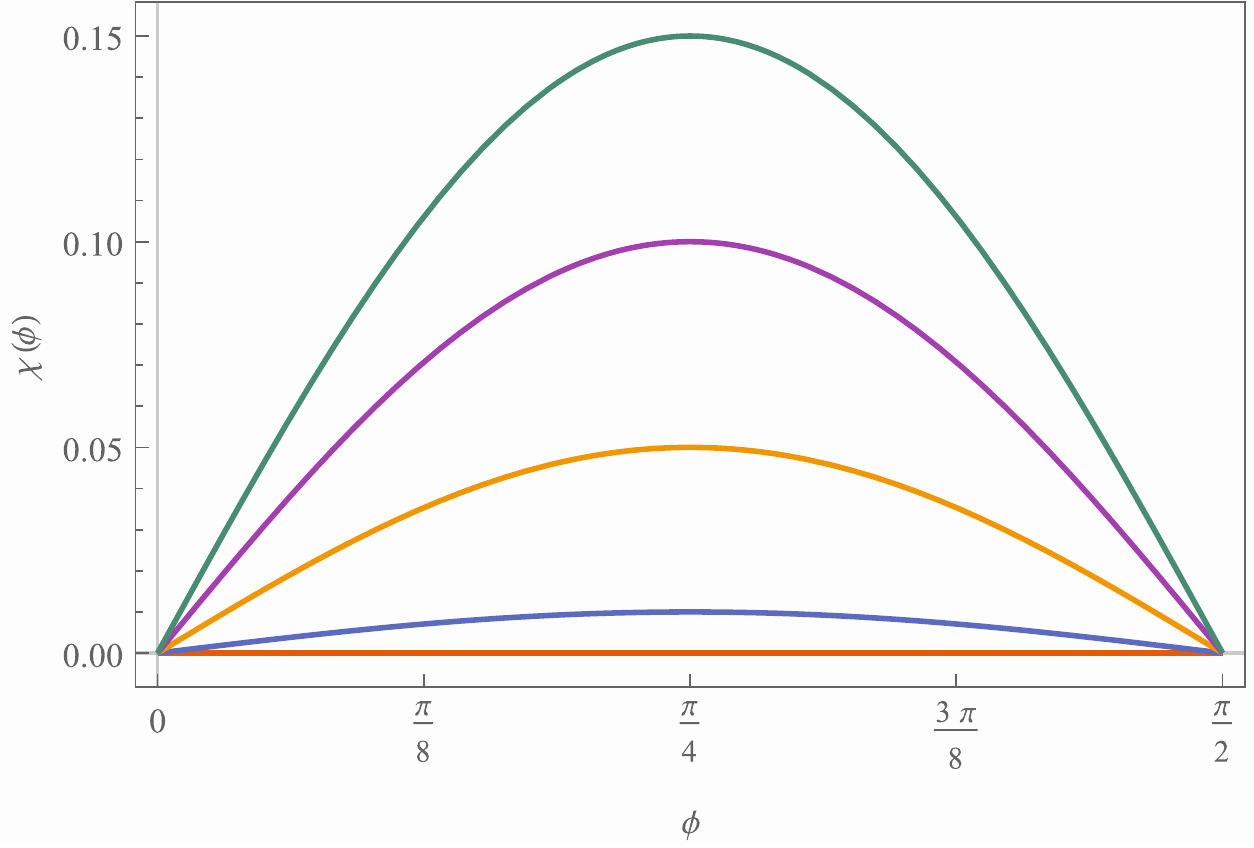}{0.6}{Form of the off-diagonal SO(2,1) breaking $dr d\phi$ term in the boundary metric $\chi(\phi)$ }{fig:chi}
\ssubfigure{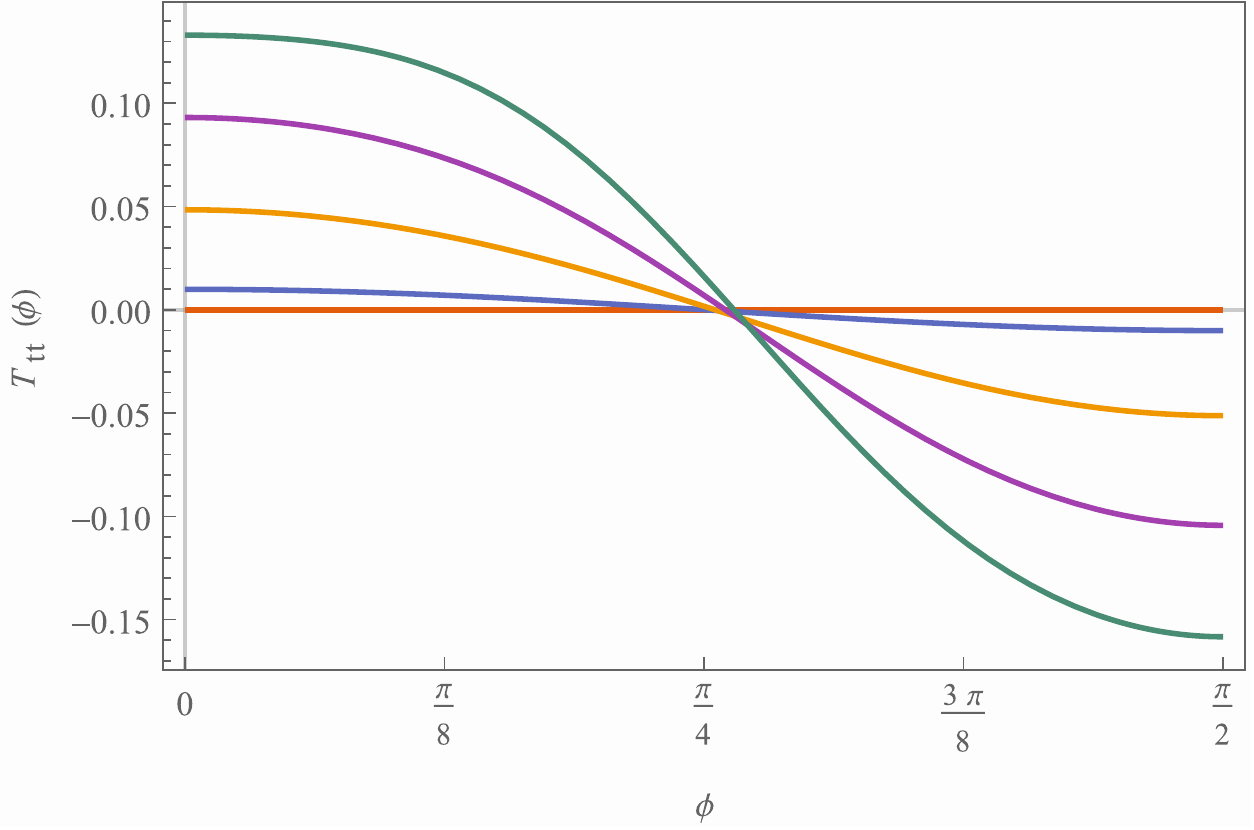}{0.6}{Resulting energy density $T_{tt}(\phi)$ for $\alpha =1$}{fig:edal0}
\ssubfigure{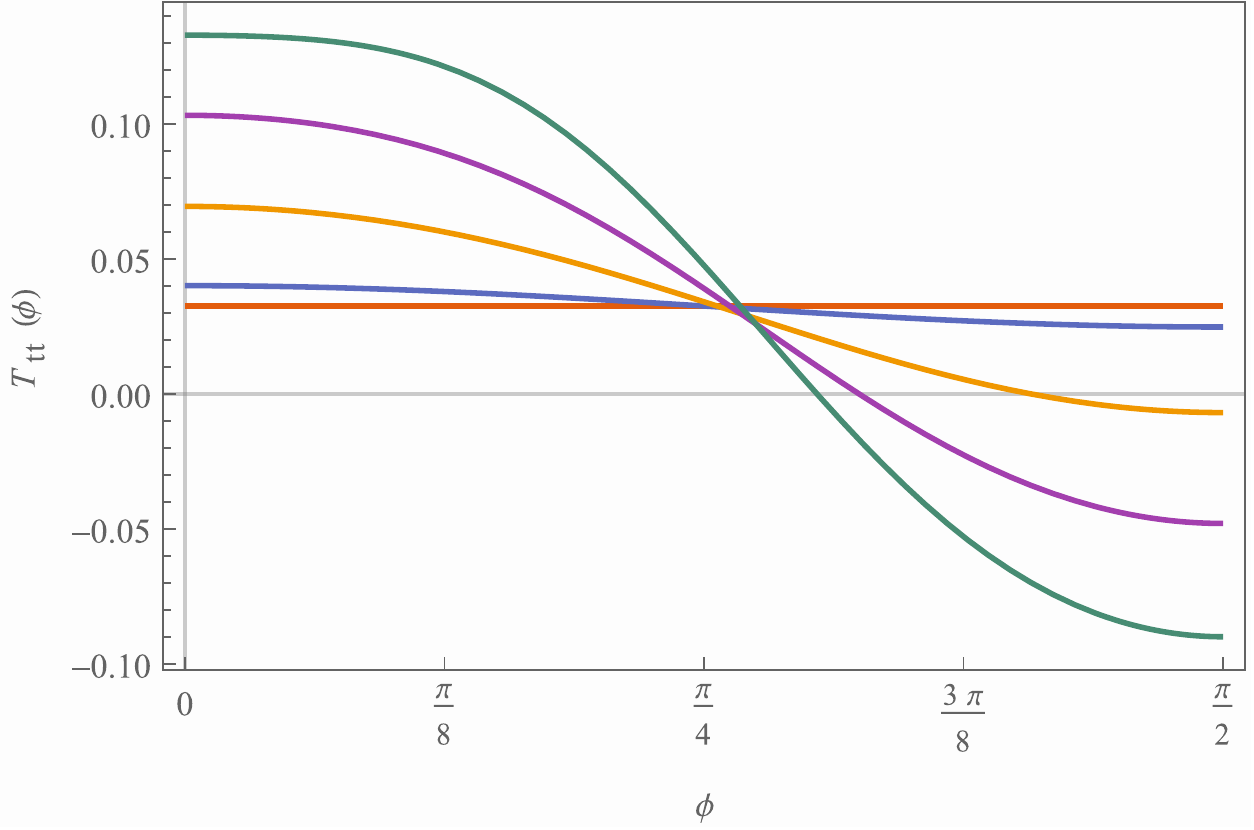}{0.4}{Resulting energy density $T_{tt}(\phi)$ for $\alpha = \frac{\sqrt{5}}{2}$}{fig:edalp}
\ssubfigure{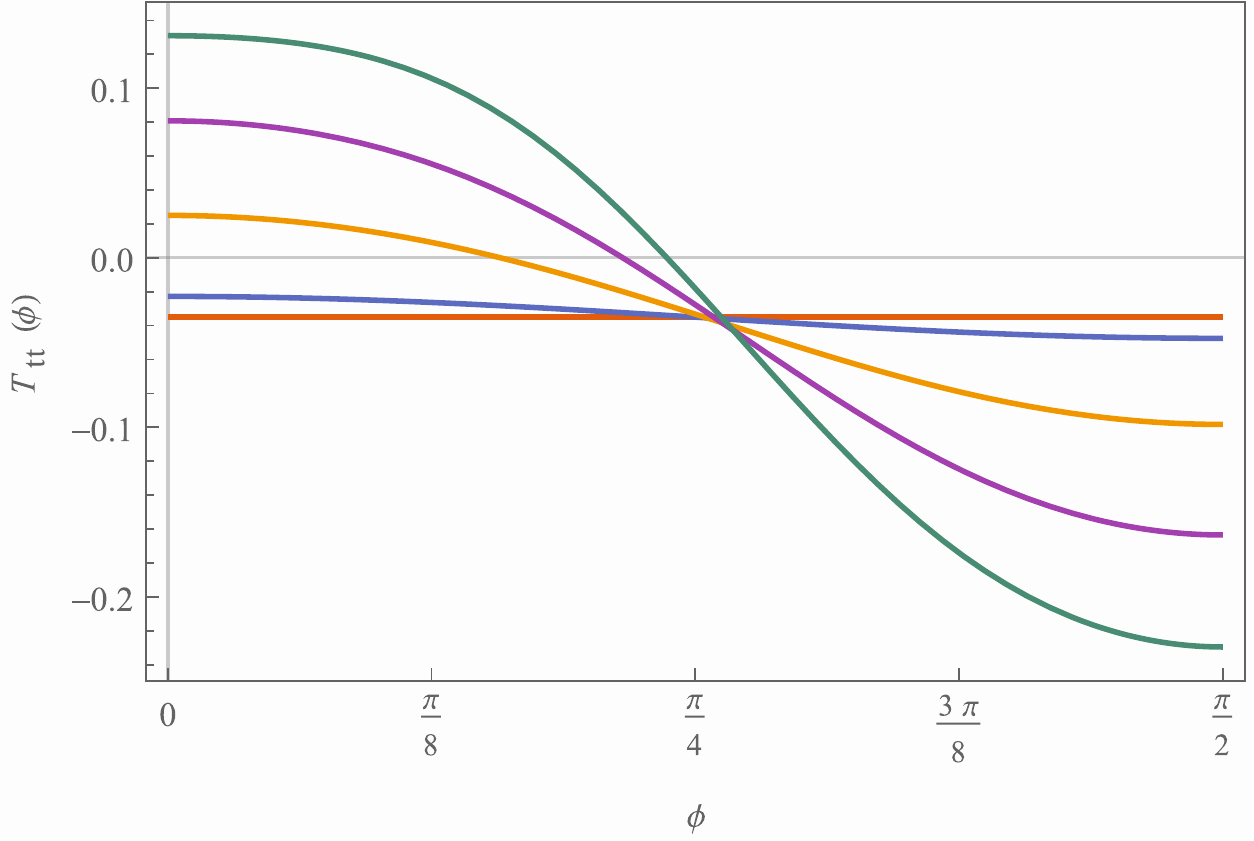}{0.4}{Resulting energy density $T_{tt}(\phi)$ for $\alpha = \sqrt{\frac{5}{6}}$}{fig:edalp}
\caption{Large scale energy density for a boundary metric with $\chi(\phi) = \frac{\lambda \sin 2\phi}{r}$, for $\lambda = \left\{0,0.01,0.05,0.1,0.15\right\}$, with 3 different conical deficits $\alpha= \left\{\sqrt{\frac{5}{6}},1,\frac{\sqrt{5}}{2}\right\}$. The energy densities go as $1/r^3$, and we've evaluated them at $r=1$. Now that we've deformed the boundary metric, the energy density is no longer positive definite. This is not a feature that is peculiar to our solutions as it was already the case in the near horizon solutions that some had positive energy density, and others negative. Roughly speaking the energy density follows the gradient of $\chi(\phi)$, and it always intersects the near-horizon energy density for the corresponding value of $\alpha$ close to when $\chi\prime(\phi)=0$. }
\label{fig:energydensity}
\end{figure}

Once we  break the $SO(2,1)$ symmetry by introducing $\chi(\phi) = \lambda \sin 2 \phi$, this becomes non-isotropic and there can be regions of positive and regions of negative energy density. In Figure \ref{fig:energydensity} we give some examples of energy density profiles for a given set of $\chi(\phi)$ profiles. The energy densities are shown at three different values of $\alpha$, one corresponding to no-conical deficit ($\alpha=1$), one to positive conical deficit, and one to negative conical deficit. Qualitatively, the behaviour doesn't change too much between the three cases. The energy density we get oscillates about the $\xi(\phi)=0$, near-horizon, value with an amplitude that grows as $\xi$ grows. The offset from the near-horizon value is largest when the gradient of $\xi^\prime(\phi)$ is largest, and vanishes close to the point where $\xi^\prime(\phi)=0$.

\section{The Singularity}
\label{sec:singularity}
To understand the nature of the singularity that replaces the extremal horizon, we examine uncharged geodesics that approach this surface. Moving, for the moment, to general dimensions $d$, we choose coordinates where the metric takes the form 
\begin{equation}
\label{eq:kkanzatz}
ds^2 = \psi(x)^2\frac{-r^4 dt^2+dr^2}{r^2} + 2 \psi(x)^2 A_i(x) d x^i \frac{dr}{r} +(h_{ij}(x)+\psi(x)^2 A_i(x) A_j(x))dx^i dx^j,
\end{equation}
so that the null surface of interest is now at $r=0$. The $x^i$ coordinates describe a $d-2$ dimensional subspace, on which $h_{ij}(x)$ is a non-degenerate metric, and $A_i(x)$ and $\psi(x)$ are a one-form and scalar respectively. The near horizon solutions have $A_i(x)=0$ when written in this form, and the generalized solutions we found above have no-zero $A_i(x)$ because they break the SO(2,1) symmetry. Using the remaining global scaling and time translation symmetries, we can integrate some of the geodesic equations using Noether's theorem. The equations we are left with are, 
\begin{equation}
 \label{eq:geodesic}
 \begin{split}
  E &= \psi^2 r^2 \dot{t} \\
  K &= E t+ \psi^2 \left(\frac{\dot{r}}{r}+A_i \dot{x}^i\right) \\
  \ddot{x}^i + \Gamma[h]^{i}_{\phantom{i}jk} \dot{x}^j\dot{x}^k &= -\frac{d \psi^i}{\psi}\left(k^2 + h_{kl}\dot{x}^k \dot{x}^l\right) + E A^i \dot{t} - F^{i}_{\phantom{i}j}\dot{x}^j\left(K-Et\right),
  \end{split}
 \end{equation}
where a dot represents differentiation with respect to an affine parameter $\lambda$. Indices are raised and lowered using $h_{ij}$, and we have defined $F_{ij} = A_{i,j}-A_{j,i}$. $K$ and $E$ are conserved quantities due to the static and scaling symmetry, and the third equation is the geodesic equation in the $x^i$ direction. The parameter $k^2$ is $0$ for null geodesics, and can be set to $1$ for timelike and $-1$ for spacelike geodesics.

By dotting the third equation in with $\dot{x}^i$ we can show that\footnote{This becomes an extra conservation law when $A_i=0$, which is because the scaling/static symmetry gets enhanced to $SO(2,1)$ in this case.}
\begin{equation}
 \frac{d}{d\lambda} \left(\psi^2(k^2 +\dot{x}_j \dot{x}^j )\right) = \frac{2 E^2 \dot{x}^i A_i}{r^2}.
\end{equation}
Substituting this equation into the conservation equations, we find the differential equation
\begin{equation}
 \frac{K}{E} = t+ \frac{1}{r^3}\frac{dr}{dt}+ \frac{1}{2 E^2} \frac{d}{dt}\left(\psi^2(k^2 + \dot{x}_j \dot{x}^j) \right),
\end{equation}
which can be solved to give
\begin{equation}
 r^2(t) = \frac{E^2}{(Et-K)^2 + C + \psi^2(k^2 + \dot{x}_j \dot{x}^j)}.
\end{equation}
The parameter $C$ is a constant of integration. From this equation it can be seen that any time-like or null geodesic will reach $r=0$ as $t \rightarrow \infty$. Using this result, we can rewrite the third equation as
\begin{equation}
   \label{eq:geox}
   \ddot{x}^i + \Gamma[h]^{i}_{\phantom{i}jk} \dot{x}^j\dot{x}^k = \left(A^i-\frac{d \psi^i}{\psi}\right)\left(k^2 + h_{kl}\dot{x}^k \dot{x}^l\right) + \frac{A^i}{\psi^2} \left((Et-K)^2 + C\right) + F^{i}_{\phantom{i}j}\dot{x}^j\left(Et-K\right).
\end{equation}

 Now, assume there exists some point on the $n-2$ dimensional $x$ manifold, $x_0$, where $A_i=d\psi_i=0$. Such a point exists, for instance, at the point $X=0$ in the solutions we constructed, because of the reflection symmetry we imposed. We can then consider freely falling particles at this point with $\dot{x}^i=0$. In this case, the particles stay at $x=x_0$. Timelike geodesics satisfy
\begin{equation}
\begin{split}
t(\lambda) &= -\frac{\cot \lambda}{B}+t_0 \\
r(\lambda)&= -B \sin \lambda,
\end{split}
\end{equation}
$t_0$ and $B$ being constant, while null geodesics are given by
\begin{equation}
\begin{split}
t(\lambda) &= -\frac{1}{\lambda}+t_0 \\
r(\lambda)&= -\lambda,
\end{split}
\end{equation} 
where in both cases we have chosen $\lambda = 0$ to be the time when the geodesic hits the surface.

Define $V^a$ to be the tangent vector $\frac{d}{d\lambda}$. The geodesic deviation tensor in the $n-2$ dimensional $x^i$ space is then given by $\tilde{R}_{ij} = R_{a i b j} V^a V^b$. This means that if the vector $X^i$ describes the infinitesimal displacement between a pair of geodesics both initially parallel to $V^a$,
\begin{equation}
V.\nabla(V.\nabla X^i)= - h^{ij}\tilde{R}_{jk}X^k.
\end{equation}
For both timelike and null cases we find that this goes as
\begin{equation}
\tilde{R}_{ij} \sim \frac{1}{\lambda^2}\left(\psi^4 F_{ik}F_{jl}\gamma^{kl} - \psi^2 \nabla_{( i}A_{j)}\right) .
\end{equation}
We therefore see that, at least for this geodesic, at $A_i(x)=d\psi(x)=0$, the geodesic deviation diverges as you approach the surface.

More generally, the geodesics seem to be very poorly behaved. As  $r\to 0$, we  have $t \rightarrow \infty$. By examining (\ref{eq:geox}), it seems to be difficult to have this, and not have divergent $\dot{x}$, unless $A^i=0$ at the point the geodesic hits $r=0$, although we note that we don't have a proof of this.

\begin{figure}
\centering
\ssubfigure{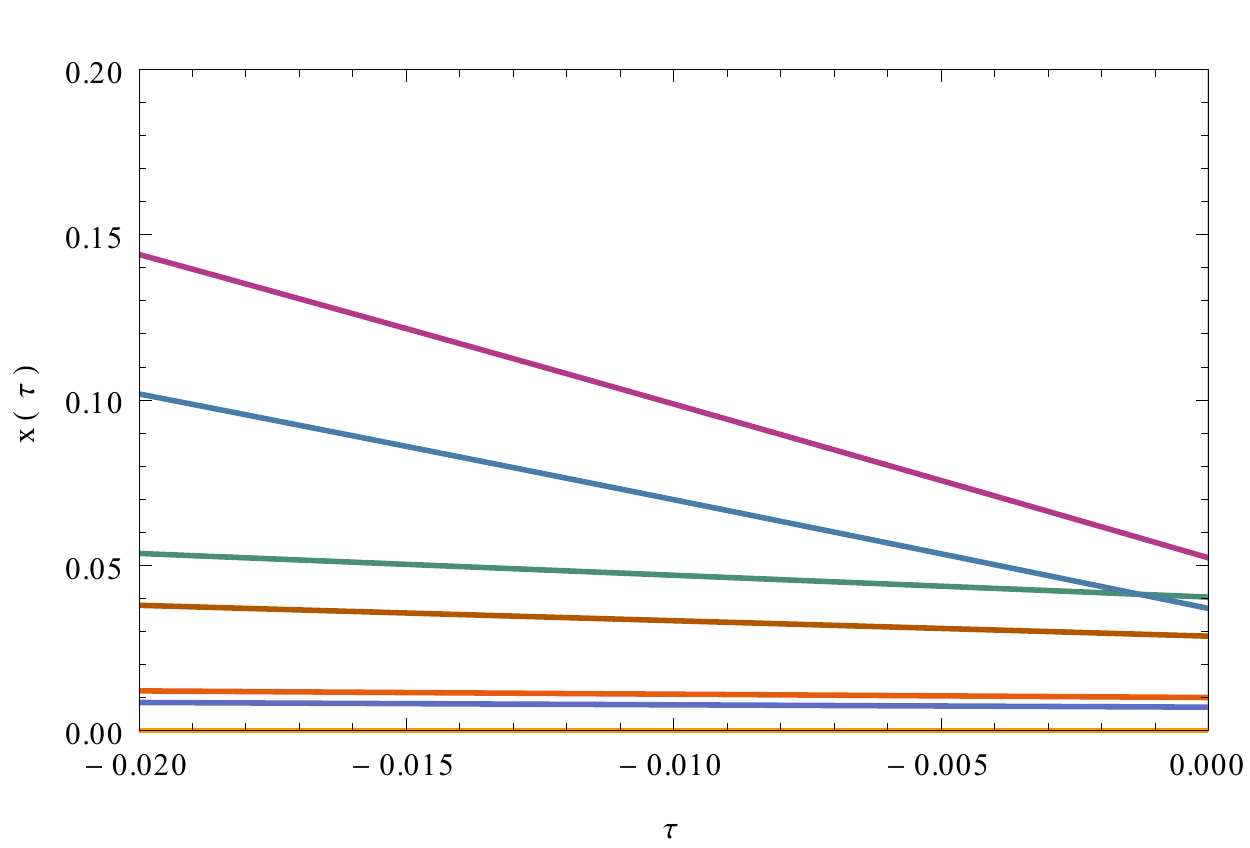}{0.4}{$AdS_4$}{fig:rt1}
\ssubfigure{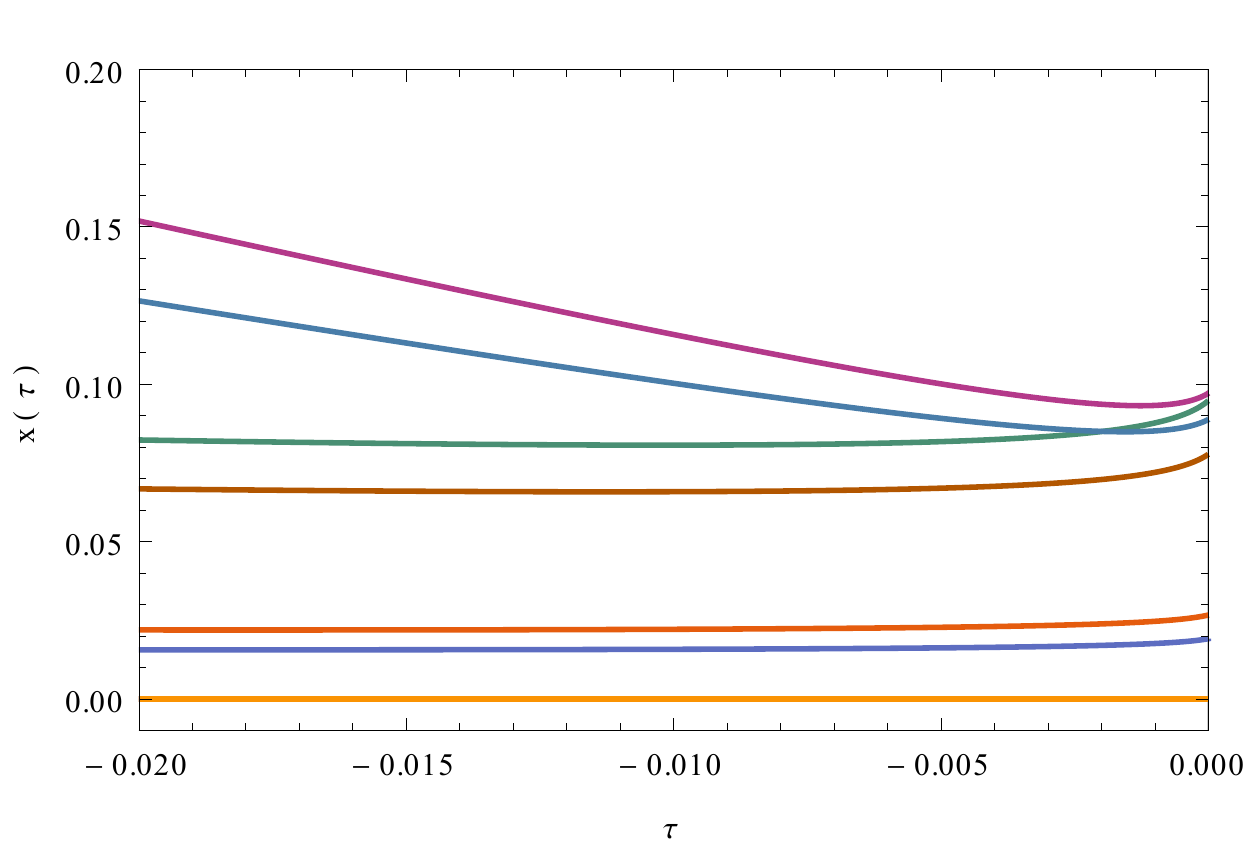}{0.4}{SO(2,1) breaking solution $\alpha =1$, $\lambda = 0.15$}{fig:rt3}
\ssubfigure{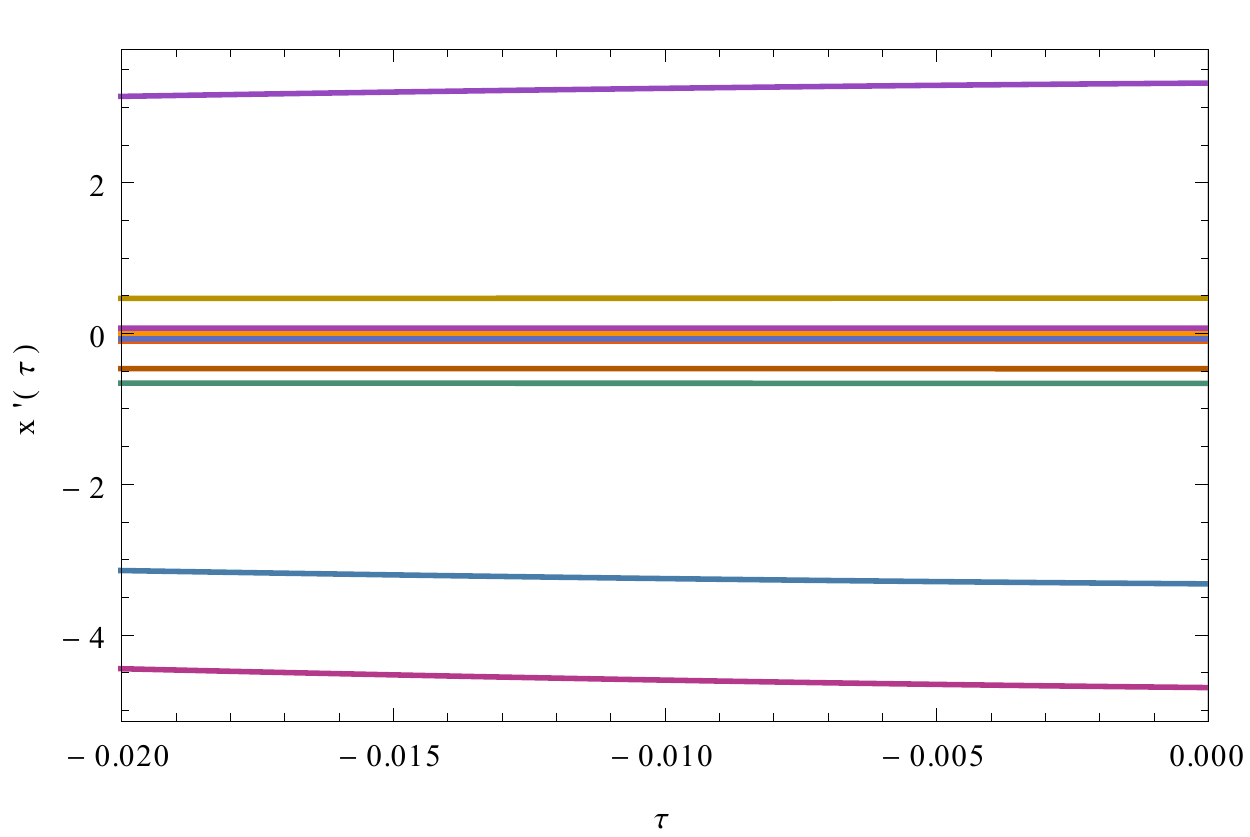}{0.4}{$AdS_4$}{fig:rt1}
\ssubfigure{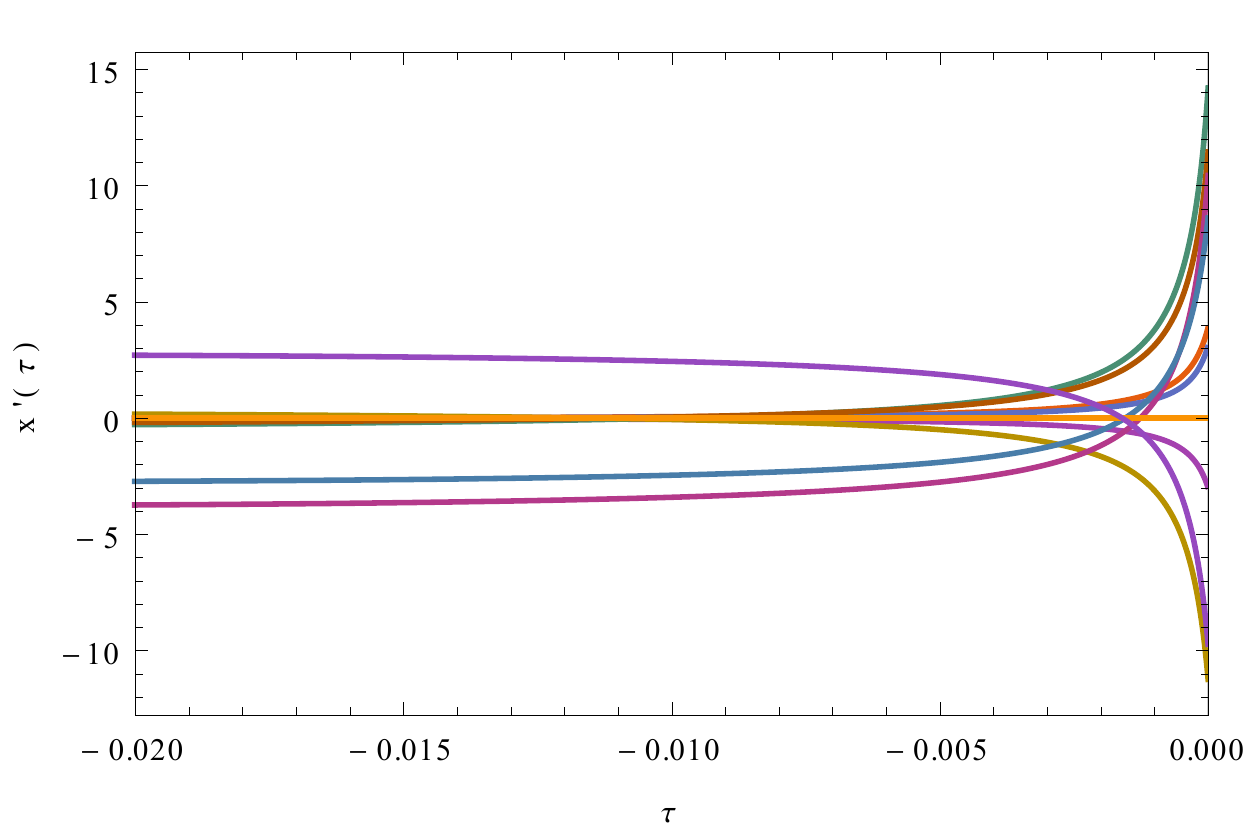}{0.4}{SO(2,1) breaking solution}{fig:rt3}
\caption{$x(\tau)$ and $x'(\tau)$ for various numerical timelike geodesics that head towards the $r=0$ surface. The proper time coordinate has been chosen so that it is $0$ when they hit the surface. As they approach the surface, nothing happens in the case of the $AdS_4$ and they just cross over a smooth extremal horizon. In the other case, all the geodesics start to swerve off as they approach the horizon, except a single geodesic that we've placed at the point $x=y=0$ with no transverse velocity ($x^\prime = y^\prime = 0$). This is the point where $A_i = d\psi_i =0$ as discussed in the text.   }
\label{fig:geodesicsx}
\end{figure}

We can, however, provide evidence for this numerically. We can take our numerically found bulk geometries (with no gauge-field), and shoot timelike geodesics towards the singularity. In Figure \ref{fig:geodesicsx} we plot $x(\tau)$, and $x'(\tau)$ for a selection of geodesics on pure-AdS and on one of our $SO(2,1)$ symmetry breaking solutions. The coordinate $x$ is one of a pair of Cartesian coordinates on our unit disk, and $\tau$ is proper time. These geodesics were evolved using an adaptive step size Runge-Kutta method, and we integrated until the step-size became too small to continue. From this we can see that, in the non-AdS case, as the geodesics get close to the singular surface there is a large transverse acceleration, and it looks like geodesics will have divergent $\dot{x}$ by the time they hit the surface. It seems reasonable to conclude, based on examination of \eqref{eq:geox} above, that the large transverse acceleration observed in the twisted case just before the end will continue and lead to a divergent $\dot{x}$.

\section{Summary}

We have found 3+1 dimensional static, scale-invariant solutions to Einstein-Maxwell equations outside the class of the known analytic solutions which correspond to the possible near horizon geometries of extremal horizons. We would argue that these are the appropriate geometries to describe the infra-red when we are considering states in holographic field theories on 2+1 dimensional boundary geometries, which, on a large scale, do not have the full SO(2,1) isometry needed to produce an extremal horizon in the bulk. We would expect that if you took these non-scale-invariant geometries, and took a large scale limit, you would end up with the scale-invariant geometries we found here. 

For instance, we can consider introducing a source electric potential into a holographic CFT which has some arbitrary form, but falls off as $A\sim \frac{dt}{r}$.
As remarked in \cite{Horowitz2014}, we can use extremal horizons to describe  such electric-potentials which asymptotically fall off as
\begin{equation}
\label{eq:nearhorizonchem}
 A= \frac{e dt }{r}.
\end{equation}
However, if the fall off breaks this axial-symmetry, 
\begin{equation}
A = \frac{V(\theta)dt }{r},
\end{equation}
the $SO(2,1)$ symmetry is broken, and we expect the singular solutions we've found to be the relevant infra-red geometries. In particular, we would expect there to be field theory states where the expectation value of the charge density falls off as $\rho(\phi)/r^2$, where the $\rho(\phi)$ has the form determined by the scale-invariant geometries we've found here. Similarly we'd expect them to have an energy density that has a $1/r^3$ fall-off again determined by these scale-invariant geometries. Note that, for the near-horizon solutions, there is no such boundary charge density away from the origin, $\rho(\phi)=0$. 

These charge densities and energy densities are the ones we gave examples of in Figures \ref{fig:largescaleresponsea0} and \ref{fig:largescaleresponsea5} (and also in Figure \ref{fig:energydensity} for the case of metric deformations). We also looked at a delta function charge at the origin of our scale-invariant states in Figure \ref{fig:origincharge}. We would expect this in usual circumstances to correspond to the overall charge of the non-scale-invariant states of which these are the large scale limit. Indeed, it is shown in \cite{Horowitz2014}, that if you start from no chemical potential in a CFT, and smoothly turn a chemical potential on with an arbitrary dependence on $r$, then it is only a marginal chemical potential which leads to a finite net contribution to the total charge. This marginal chemical potential is the scale-invariant $A \sim V dt/r$ which we have in our singular geometries.

One limitation we ran into was the lack of smoothness in our coordinate choice at the boundary. This is explained in Appendix \ref{sec:logarithm}, and it stems from our choice of reference metric. An interesting question for future work using the technique of the harmonic Einstein equations would be how to pick a reference metric that leads to coordinates that are as smooth as possible.

\section*{Acknowledgements}
We would like to thank Toby Wiseman for many helpful discussions, advice, and feedback. The numerical work was carried out using the facilities of the Imperial College High Performance Computing Service. All plots were produced using Mathematica. AH is supported by an STFC studentship.

\appendix

\section{Gauge Fixing Static Scale Invariant Geometries}
\label{sec:scaleinvapp}
Here we explain why the most generic $2+1$ dimensional static, scale-invariant boundaries, where the scaling symmetry takes the form 
\begin{equation}
\label{eq:scalingsymmetryapp}
\begin{split}
t &\rightarrow \epsilon t \\
r &\rightarrow \epsilon r,
\end{split}
\end{equation}
can be written in the form \eqref{eq:twistedboundary}.

First we consider the metric. If we write down all the terms in the metric that satisfy the symmetries, we get
\begin{equation}
g = -A (\phi)\frac{dt^2}{r^2} + B(\phi) \frac{dr^2}{r^2} + S(\phi)^2 d\phi^2 +2 H(\phi) d\phi \frac{dr}{r}.
\end{equation}
We are left with the following coordinate freedom
\renewcommand{\theenumi}{(\roman{enumi})}
\renewcommand{\labelenumi}{\roman{enumi}}
\begin{enumerate}
\item Weyl Scaling of the metric since we only care about it's conformal class:  $g^\prime = \exp(f(\phi))g$ \label{weyl}
\item The redefinition of the $r$ coordinate:  $r = \lambda(\phi) r^\prime$ \label{rgauge}
\item The redefinition of $\phi$: $\phi = \phi(\phi^\prime)$. \label{diffeo}
\end{enumerate}

Note, that there is an additional freedom corresponding to the rescaling $t = \mu t^\prime$. However, because of the symmetry of \eqref{eq:scalingsymmetryapp}, this is equivalent to \ref{rgauge} with $\lambda(\phi) = \frac{1}{\mu}$.

We can use the freedom of \ref{weyl} and \ref{rgauge} to set $A=B=1$. This uses up these two coordinate freedoms completely. We can then use \ref{diffeo} to set
\begin{equation}
\alpha d \phi^\prime =  S(\phi) d\phi.
\end{equation}
We choose to fix $\phi$ and $\phi^\prime$ to have canonical period $2 \pi$. We are therefore left over with the constant $\alpha= \frac{1}{2\pi}\int S(\phi) d\phi$ because we are restricted to redefinitions of $\phi$ that preserve this period. 

Dropping the prime we find that we can write the most general boundary as
\begin{equation}
\label{eq:twistedboundaryapp}
g = -\frac{dt^2}{r^2} + \frac{dr^2}{r^2} + \alpha^2 d\phi^2 + 2 \alpha \chi(\phi) d\phi \frac{dr}{r}.
\end{equation}
Therefore, a 2+1 dimensional twisted cone boundary is specified by one periodic function $\chi(\phi)$ and one constant $\alpha$ which we take to be positive.

If we demand that these coordinates are non-degenerate, we require
\begin{equation}
\left \vert \chi(\phi) \right \vert < 1,
\end{equation}
so that $\det(g) < 0$. 


To check that these boundaries aren't secretly conformally flat, we look at the Cotton-York tensor. The non-vanishing components of this symmetric, traceless tensor are
\begin{equation}
\begin{split}
C^{tr}&=\frac{2 \chi (\phi ) \chi\prime(\phi )^2}{\alpha \left(\chi (\phi )^2-1\right)^3}-\frac{\chi\prime\prime(\phi )}{2 \alpha  \left(\chi (\phi )^2-1\right)^2}\\
C^{t\phi} &=-\frac{\chi\prime(\phi )}{\alpha  r \left(\chi (\phi )^2-1\right)^2}.
\end{split}
\label{eq:cottonyork}
\end{equation}
The vanishing of this tensor is the necessary and sufficient condition for conformal flatness in three dimensions. We see that for the case of constant $\chi(\phi)=\mu$ this condition is satisfied, and there must be some extra gauge freedom. In this special case, we can make the redefinition $\phi \rightarrow \phi+\mu \log r$ to get rid of $\chi(\phi)$\footnote{As is evident from Equation \ref{eq:cottonyork}, we cannot in general use a redefinition of this form to get rid of $\chi$. If we set $\phi \rightarrow \phi + \chi(\phi) \log r$ we pick up log terms in the metric unless $\chi(\phi)$ is constant.}.

The gauge field is simpler. We have a $t\rightarrow -t$ symmetry which means that it is consistent to consider a purely electric source potential, with no spatial vector potential. When combined with the scaling symmetry, this allows us to immediately write down
\begin{equation}
A = V(\phi) \frac{dt}{r}.
\end{equation}
Note that we cannot use the redefinition of $r$, \ref{rgauge}, to get rid of this, as we have already used up this freedom in fixing the metric.

\section{Convergence}
\label{sec:convergence}
In order to make the claim that the solutions we find are good approximations for actual continuum solutions, we need to have some checks of convergence. We perform these checks using \nth{6} order finite difference in the radial direction, and spectral interpolation in the angular direction, for a selection of the $a,b$ pairs we collected for the solutions with a gauge field. We also performed corresponding checks for the pure gravitational solutions and we see similar behaviour.

\begin{figure}
\subfigure[]{\includegraphics[width=\textwidth]{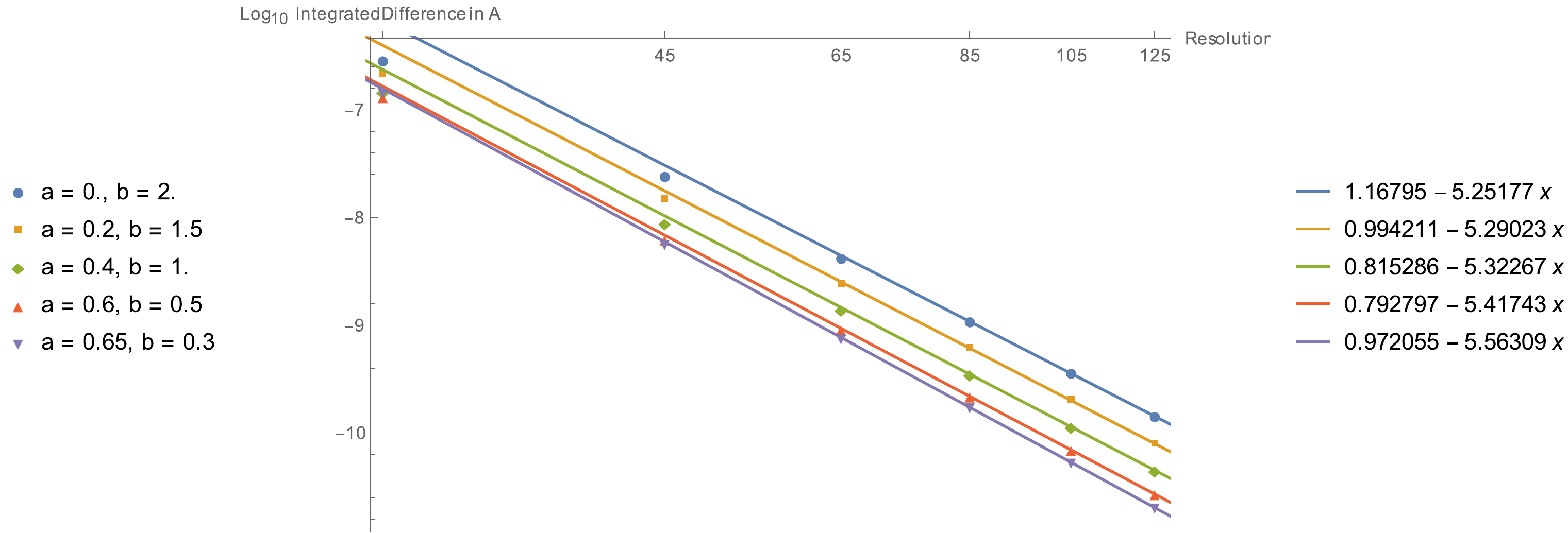}\label{sf}}
\subfigure[]{\includegraphics[width=\textwidth]{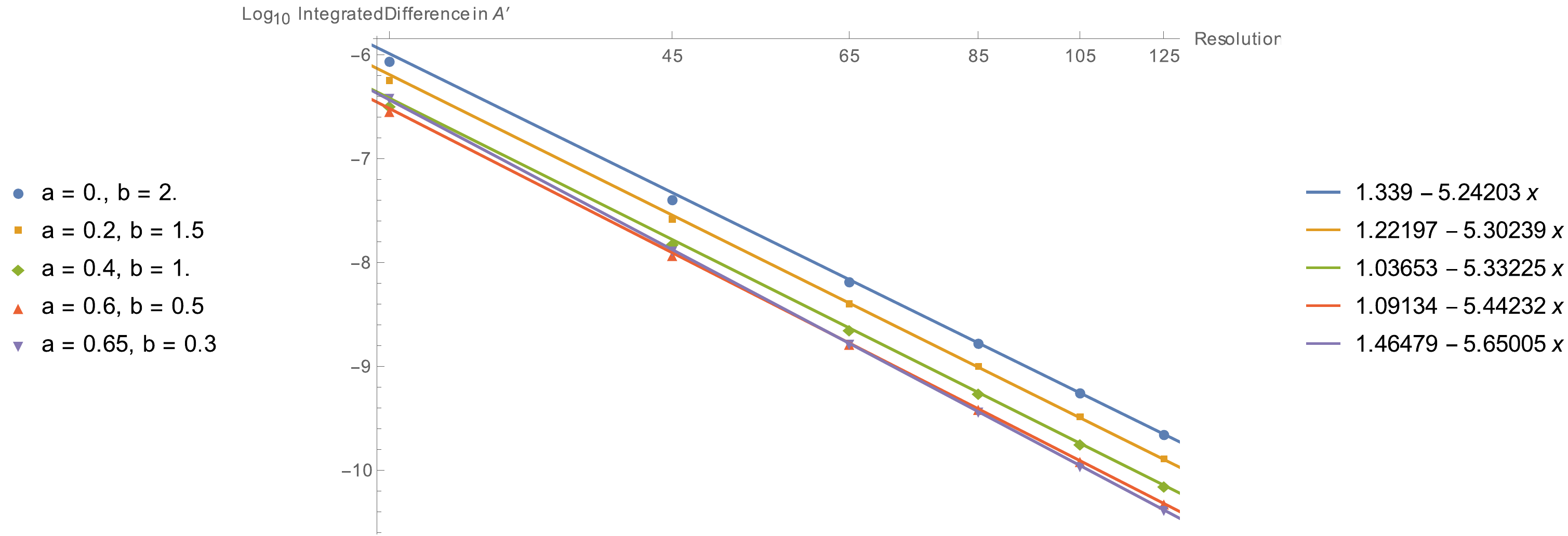}\label{sd}}
\subfigure[]{\includegraphics[width=\textwidth]{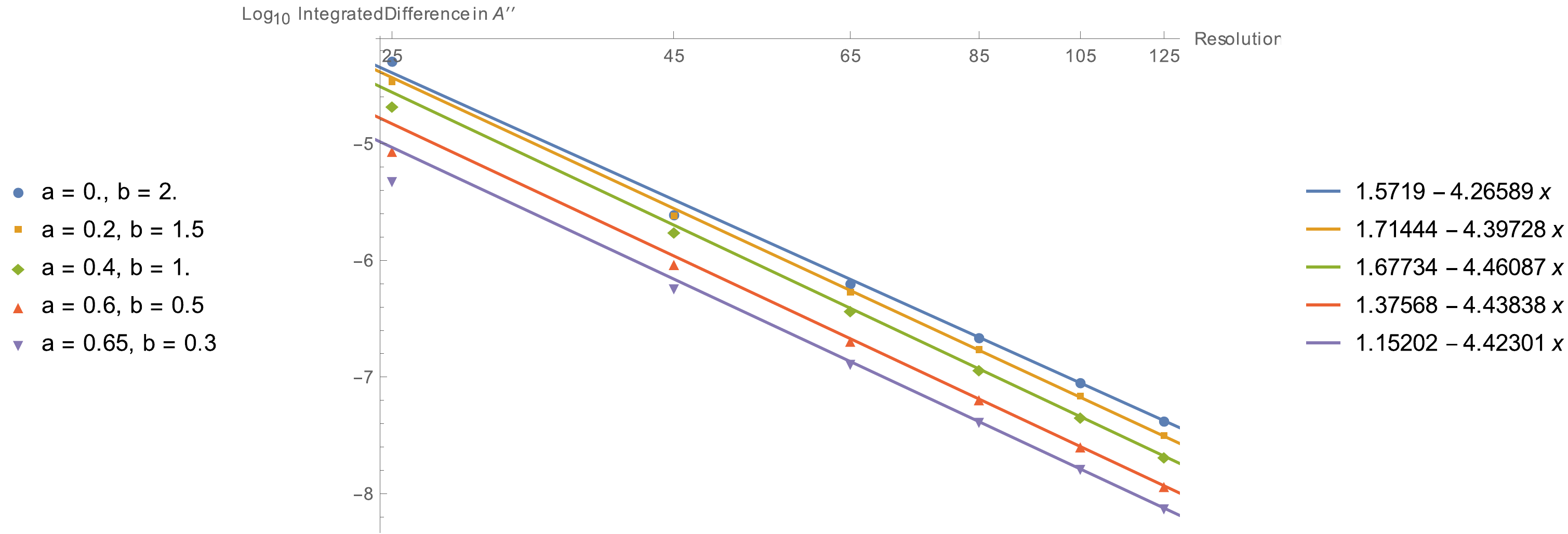}\label{sdd}}
\caption{Integrated absolute value of the shift in the solutions between subsequent $X$ resolutions with $20$ points in the $\phi$ direction, taken on the slice $\phi=\frac{\pi}{8}$. We take the case of the function $A$ and we plot the shifts in (a) it's value, (b) it's first derivative, and (c) it's second derivative. These are log-log plots, and our linear  fit to the final few points give us an estimate of the order of convergence. In these fits, $x=\log_{10}$Resolution. These seem to indicate fourth order convergence for the function and it's first derivative, and second order convergence for it's second derivative.  }
\label{fig:bulkconvergence}
\end{figure}

\begin{figure}
\subfigure[]{\includegraphics[width=0.8\textwidth]{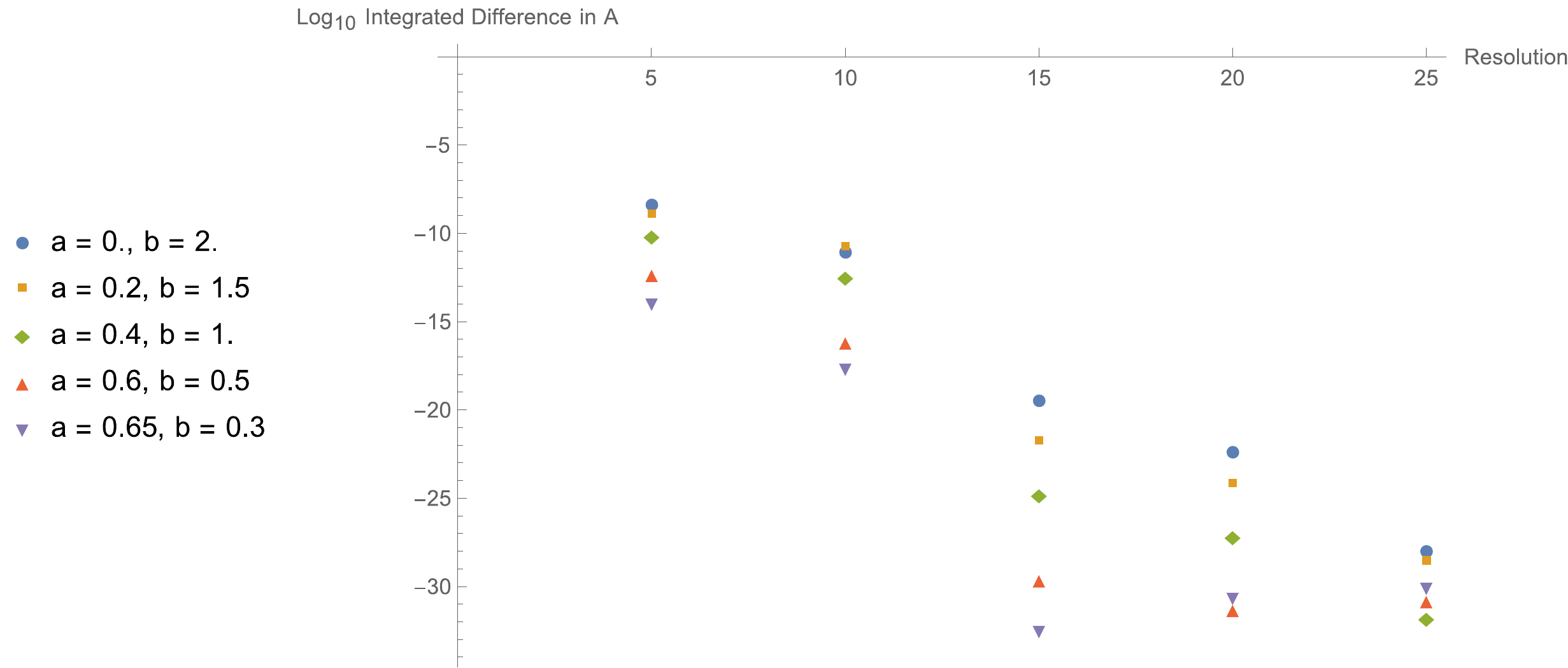}\label{sf}}
\subfigure[]{\includegraphics[width=0.8\textwidth]{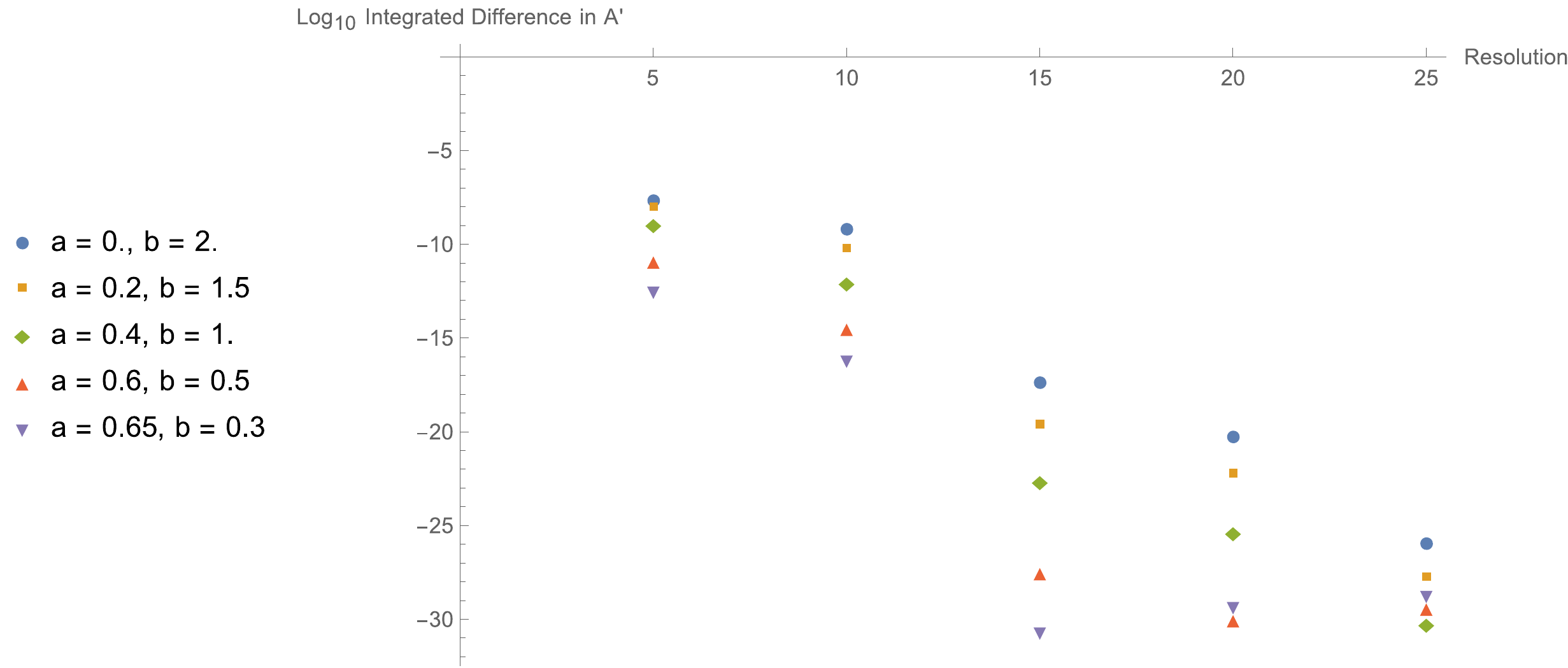}\label{sd}}
\subfigure[]{\includegraphics[width=0.8\textwidth]{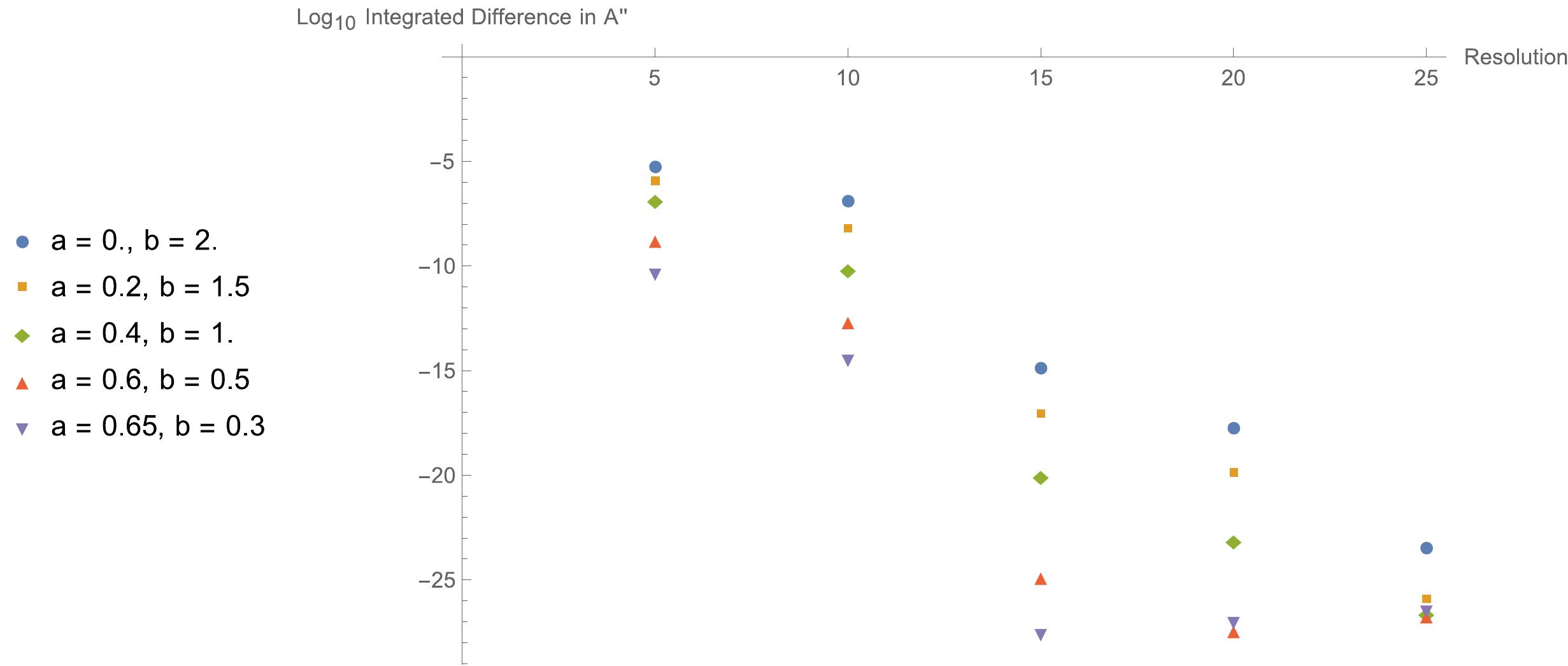}\label{sdd}}
\caption{Integrated absolute value of the shift in the solutions between subsequent $\phi$ resolutions, with $65$ points in the $X$ direction, taken on the slice $\phi=\frac{\pi}{8}$. We take the case of the function $A$ and we plot the shifts in (a) it's value, (b) it's first derivative, and (c) it's second derivative. These plots indicate approximately exponential convergence with $\phi$ resolution. When you continue above a $\phi$ resolution of $25$ the $X$ resolution becomes limiting. }
\label{fig:bulkconvergencephi}
\end{figure}

We start by checking convergence of our solutions directly. Plotted in Figure \ref{fig:bulkconvergence} is the integrated absolute value of the shift of one of the functions in our solutions between successive $X$ resolutions at a fixed $\phi$ resolution of $20$ points. Comparing these to the corresponding plots in Figure \ref{fig:bulkconvergencephi}, where we fix the $X$ resolution to 65 and vary the $\phi$ resolution, we see that's its the dependence on $X$ resolution that dominates as the shifts in this case are orders of magnitude bigger. 

We therefore focus on dependence $X$ resolution. Since we are using \nth{6} order finite difference, we might expect these shifts to fall off as $n^{-7}$ (so that the offset from the continuum values falls as $n^{-6}$), where $n$ is the number of points. However, what we instead see is an $n^{-5}$ fall off in the shifts of the function and its first derivative, and then more like $n^{-4}$ for it's second derivative. This is consistent with the limitation of smoothness at the boundary coming from our choice of coordinates, as elaborated in Appendix \ref{sec:logarithm}.  Indeed, in Figure \ref{fig:boundaryconvergence} we plot shifts in just the boundary values of the field, and we see slower convergence. The fifth derivative is not converging at all, and it is at this order that the boundary expansion in Appendix \ref{sec:logarithm} tells us the lack of smoothness gets introduced.

\begin{figure}
\subfigure[]{\includegraphics[width=\textwidth]{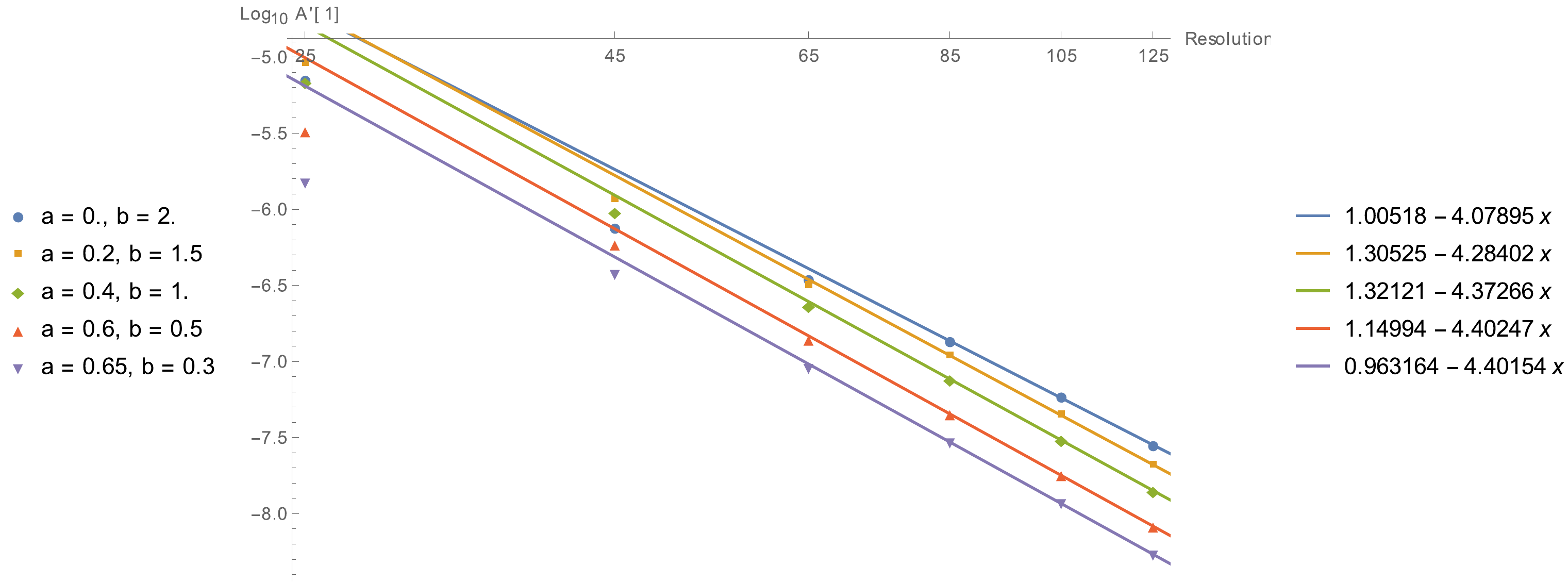}\label{sfb}}
\subfigure[]{\includegraphics[width=\textwidth]{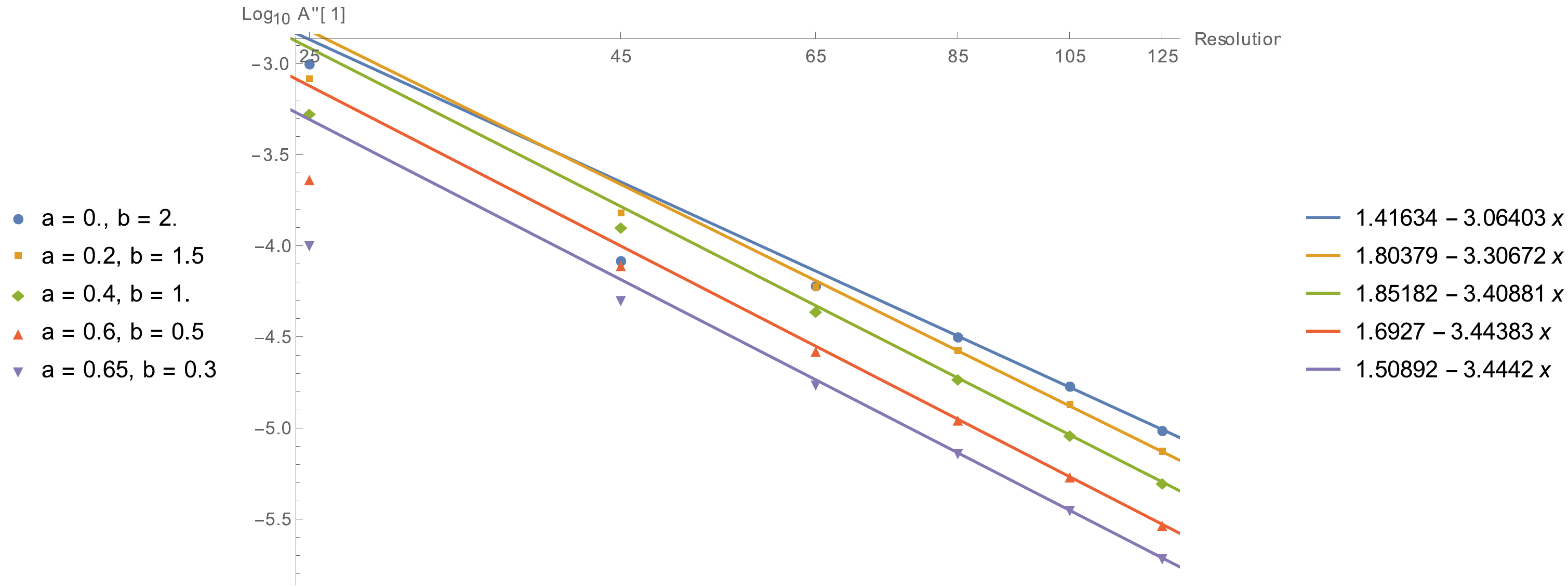}\label{sdb}}
\subfigure[]{\includegraphics[width=\textwidth]{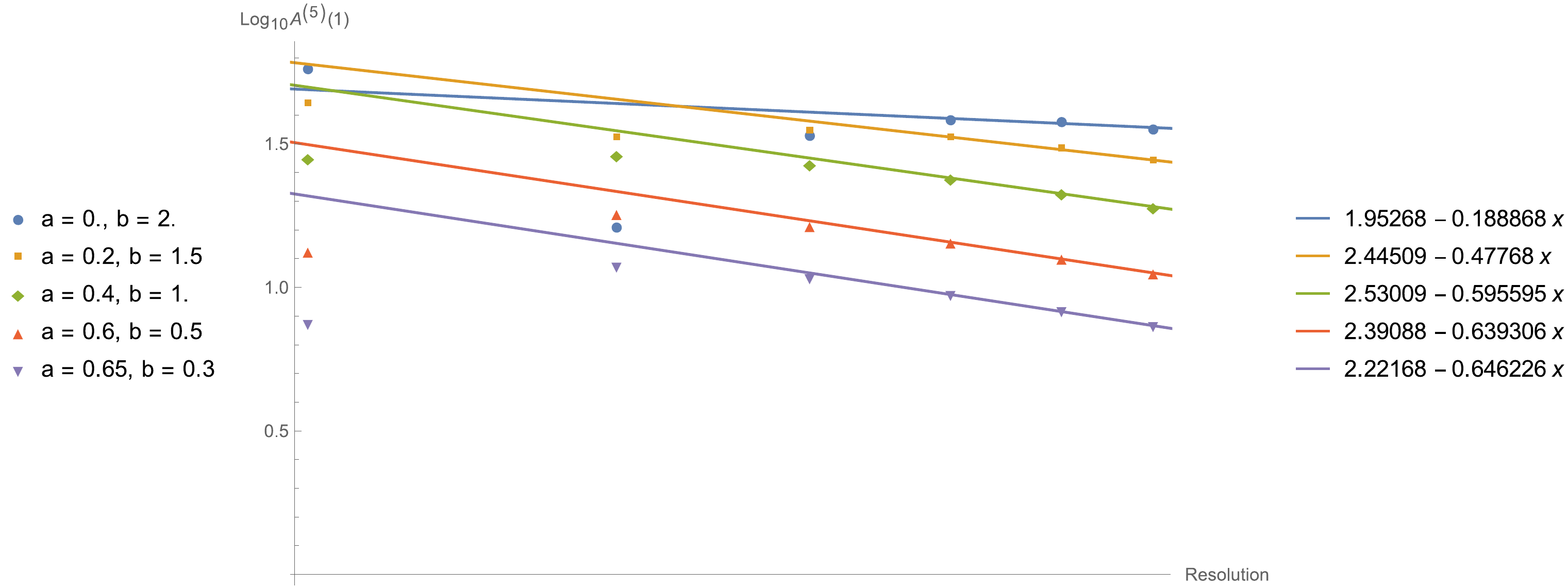}\label{sddb}}
\caption{Absolute value shift in the boundary value of the solutions between subsequent $X$ resolutions, taken at $\phi = \frac{\pi}{8}$. We take the case of the function $A$ and we plot the shifts in (a) it's first derivative, (b) it's second derivative, and (c) it's fifth derivatives. Linear fits are made to the final few points with $x=\log_{10}$Resolution. Convergence is slower than the bulk convergence in Figure \ref{fig:bulkconvergence}, and at fifth order we see the effect of the lack of smoothness explained in Appendix \ref{sec:logarithm}.}
\label{fig:boundaryconvergence}
\end{figure}

\begin{figure}
\subfigure[]{\includegraphics[width=\textwidth]{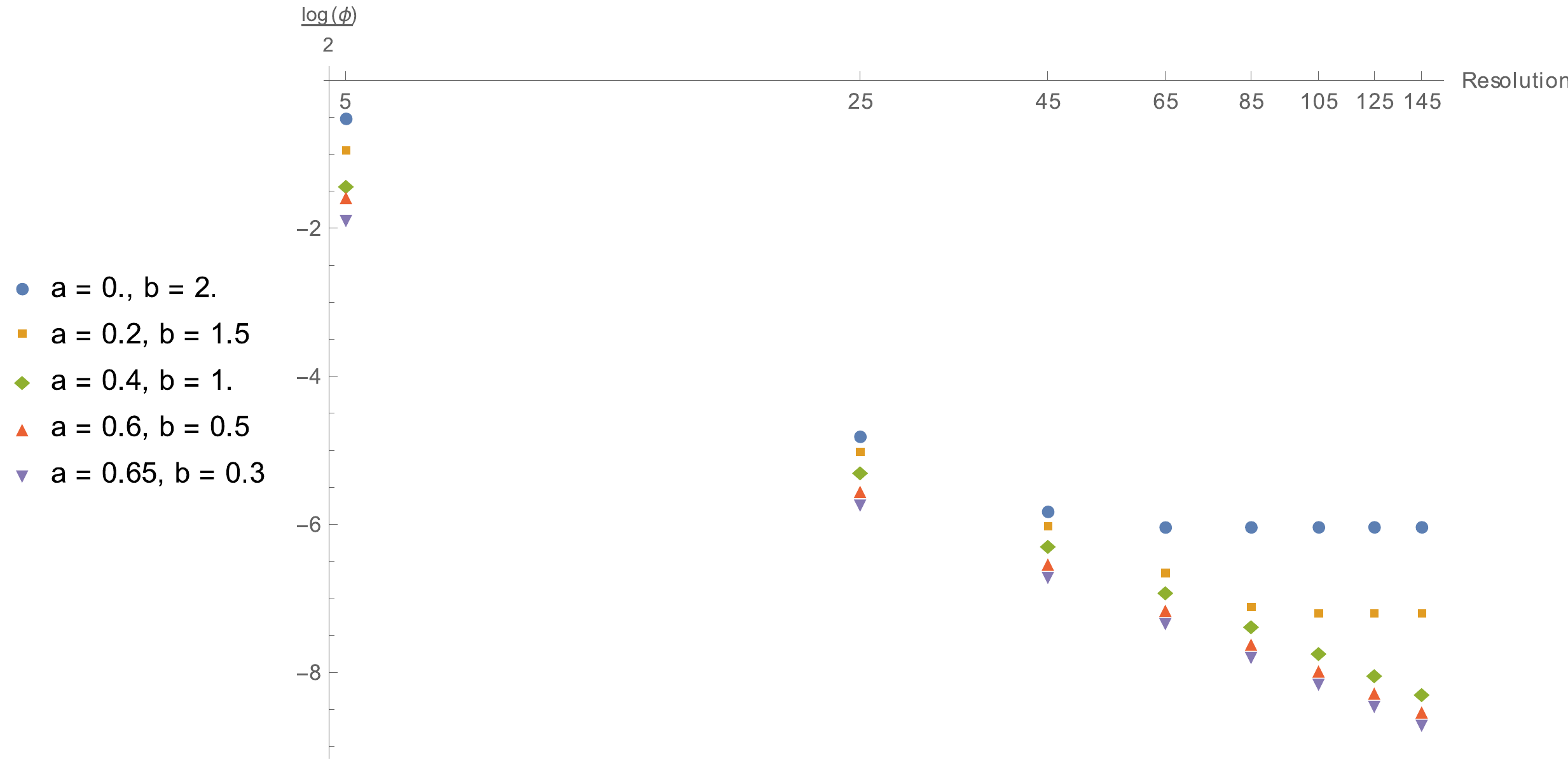}\label{sffq}}
\subfigure[]{\includegraphics[width=\textwidth]{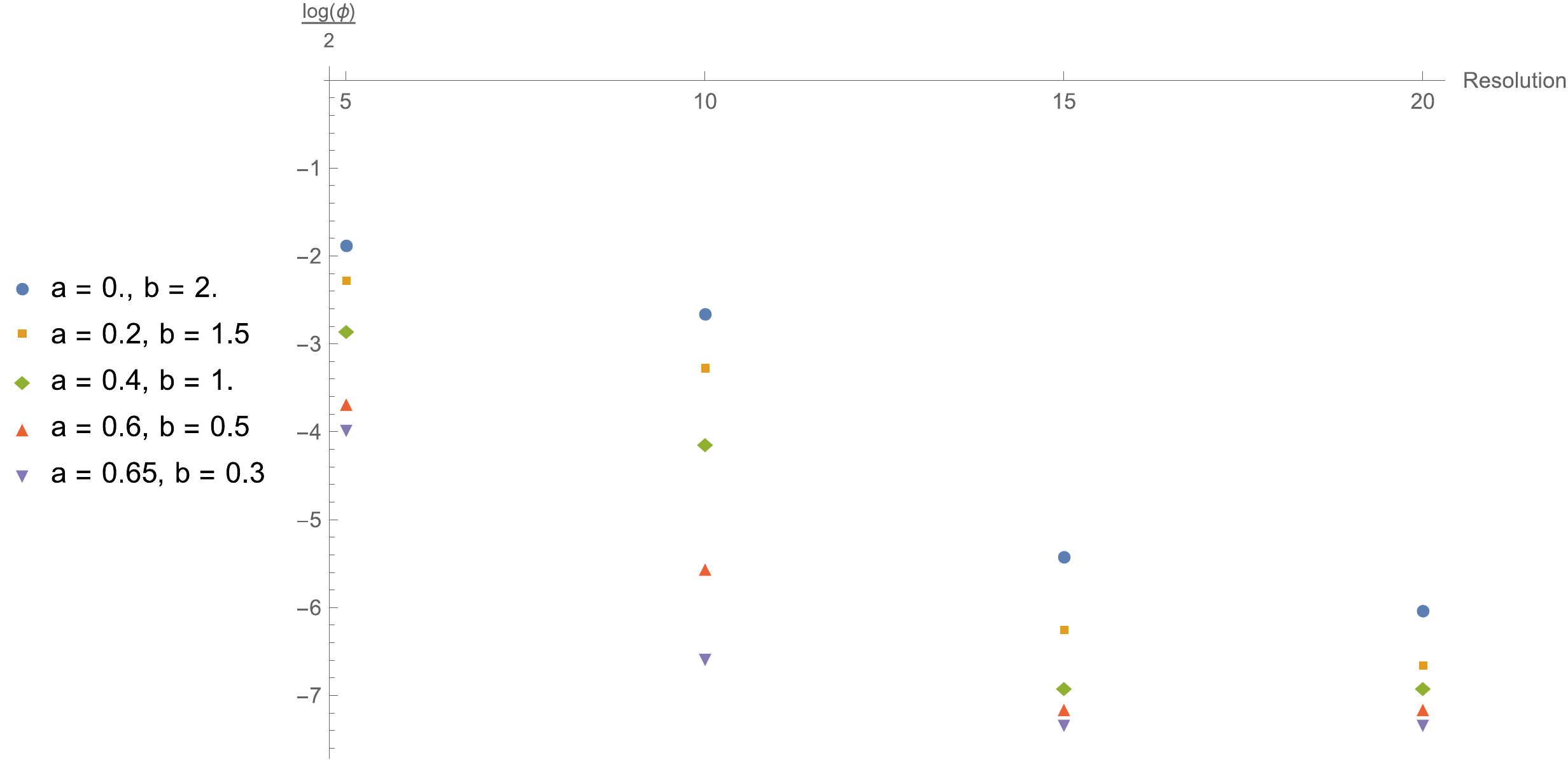}\label{sdwefb}}
\caption{The maximum value of $\phi=\xi^\mu \xi_\mu$ as a function of resolution. In (a) we see the dependence on $X$ resolution at a fixed $\phi$ resolution of $20$ points. Initially we see what looks like power law behaviour in all these solutions, but a couple of them start to level off at around $65$ points. This seems to be where the $\phi$ resolution is becoming the limiting factor. In (b) we plot the dependence on $\phi$ resolution at a fixed $X$ resolution of $65$ points. Above the $\phi$ resolution of $20$ which we plot up to the $X$ resolution becomes the limiting factor. For the parts of (a) which look like a power law, the slope is about $-3.9$, which is consistent with the third order convergence seen in the second derivatives in Figure \ref{fig:bulkconvergence}.   }
\label{fig:solitonconvergence}
\end{figure}

As a further check, we can examine our gauge condition. The vector $\xi^\mu = g^{\alpha\beta}\left(\chris{\mu}{\alpha}{\beta}-\chrisg{\mu}{\alpha}{\beta}{\bar{\Gamma}}\right)$ should be converging to zero. As in our case it's a space-like vector, it is sufficient to look at $\phi = \xi^\mu \xi_\mu$. It's dependence on both $X$ resolution and $\phi$ resolution is shown in Figure \ref{fig:solitonconvergence}. These seem to demonstrate convergence, although in the latter part of \ref{sffq}, in which we plot dependence on $X$ resolution, we are limited by the fixed $\phi$ resolution, and vice versa for \ref{sdwefb}. Since our solutions satisfy the harmonic Einstein equation, this is an indirect test of how well they satisfy the Einstein equations. We see the same behaviour when we examine the trace of the Einstein equation.

We have discussed the limitations of smoothness at the boundary, but we also need to check how smooth our fields are in the bulk\footnote{As described in Section \ref{sec:singularity}, this manifold is singular, however we can still impose smoothness away from the singularity $r=0$. This amounts to demanding that, as defined in \eqref{eq:abstractanzatz}, the scalars $S_1$ and $S_2$, the vector $\omega$, and the metric $g_2$ are smooth over the 2-d disk $\mathcal{M}$}. The convergence plots in Figure \ref{fig:bulkconvergence} give us some degree of confidence that the metric functions are at least $C^4$, but the coordinates that analysis is done with respect to are polar coordinates, so this might go wrong at the origin. To check against this, we move to coordinates that aren't degenerate at this point. 

The simplest choice of such coordinates are the Cartesian coordinates ($x$,$y$) related to these polar coordinates in the usual way. We therefore transform the scalars $S_1$, $S_2$ and $V$, the vector $\omega$, and the metric $g_2$ as defined in \eqref{eq:abstractanzatz}, into these coordinates, and examine an expansion of their components near the origin $x=y=0$. 

\begin{figure}
\subfigure[$X^0$]{\includegraphics[width=\textwidth]{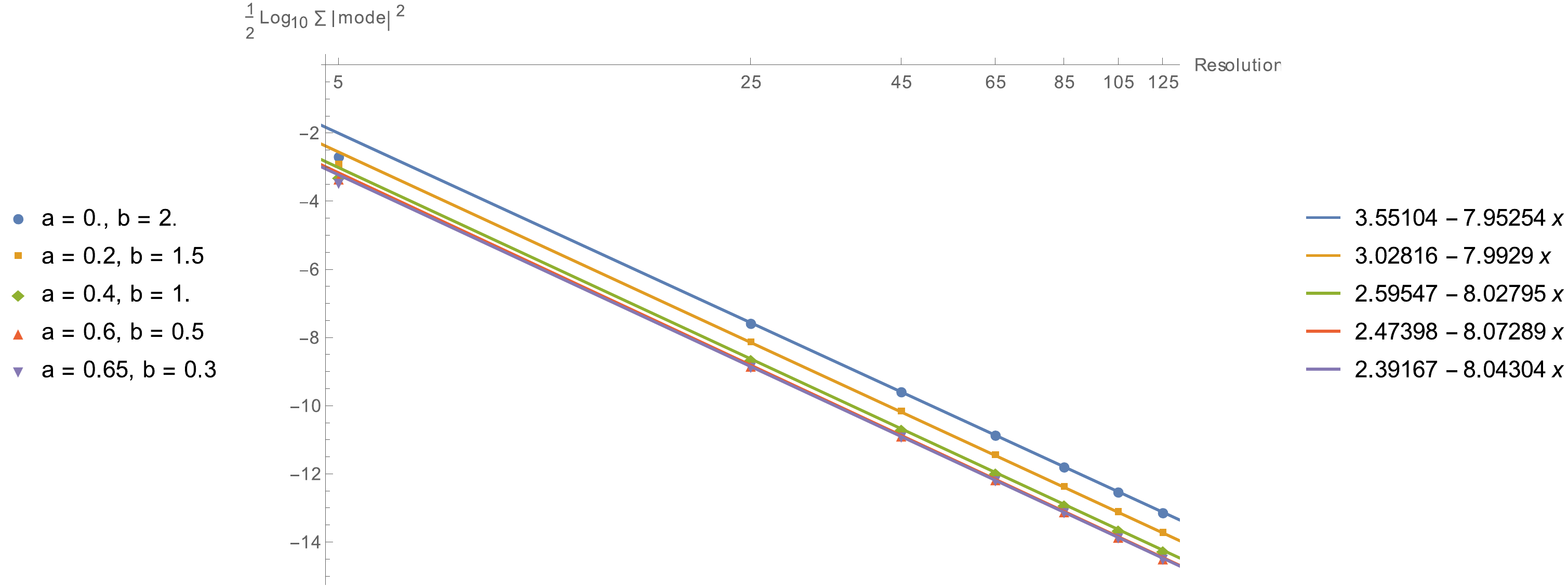}\label{sffq1}}
\subfigure[$X^1$]{\includegraphics[width=\textwidth]{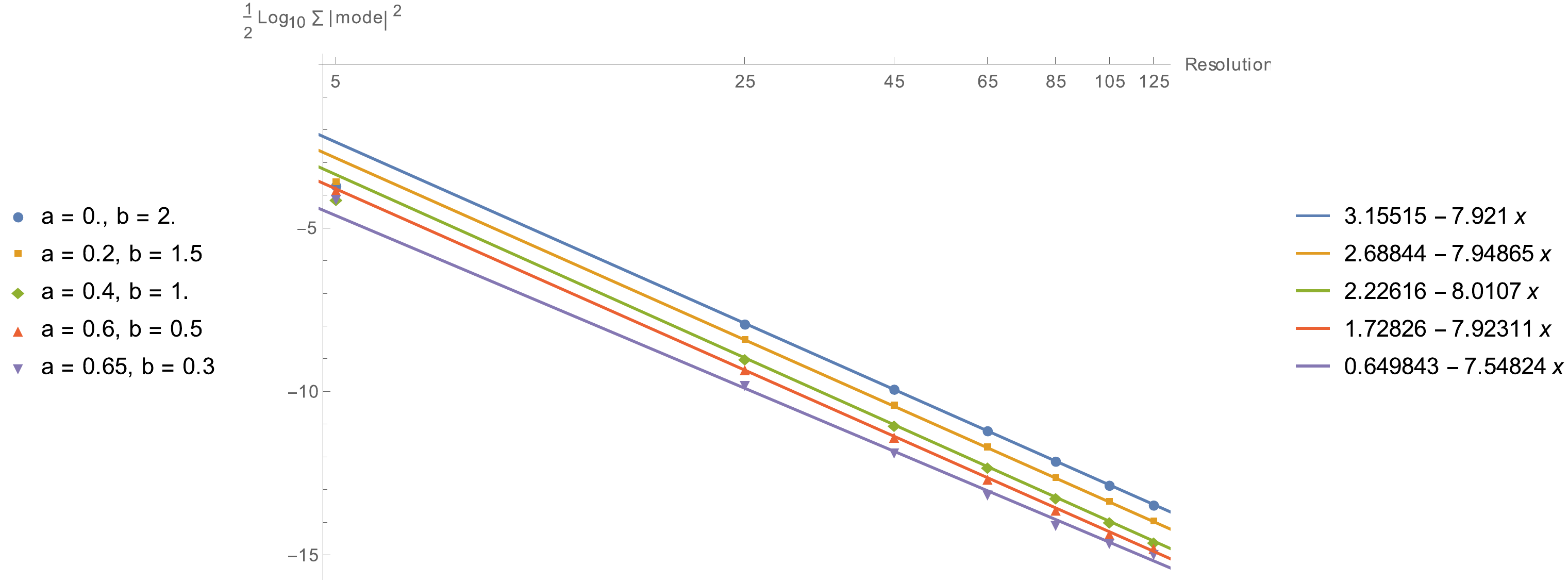}\label{sffq2}}
\subfigure[$X^4$]{\includegraphics[width=\textwidth]{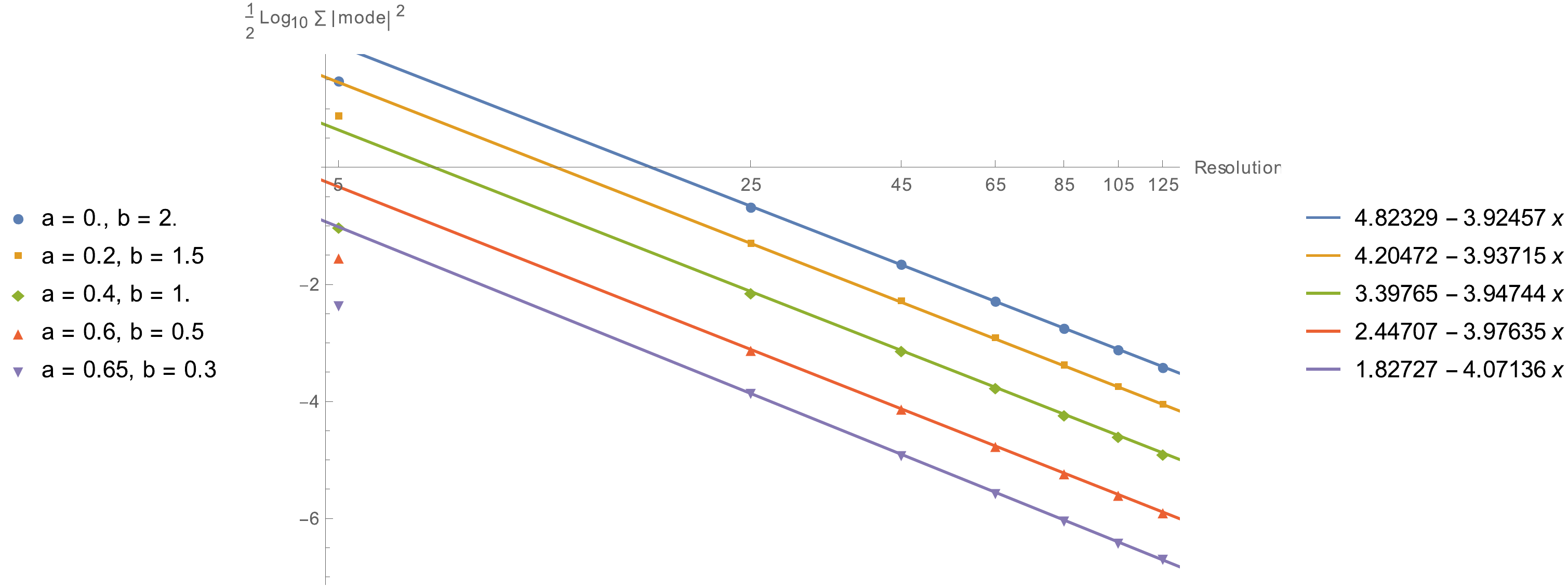}\label{sffq3}}
\caption{ We've expanded our fields near the origin $X=0$ in a Cartesian basis ($x$,$y$) and checked that derivatives in $x$ and $y$ are well defined. Thanks to our parity symmetry, this amounts simply to checking that at each order $X^n$, if we expand the $\phi$ dependence in Fourier modes, there are no modes above $\cos n \phi$ and $\sin n \phi$. We plot here the sum of the squares of the modes $\cos m \phi$ and $\sin m \phi$  at order $X^n$ for $m > n$ summed over all the components of our fields, for (a) $n =0$, (b) $n=1$ and (c) $n=4$, at a series of $X$ resolutions, at a fixed $\phi$ resolution of $15$ points. These show that the functions are well behaved at least up to fourth derivatives at the origin. }
\label{fig:originsmoothness}
\end{figure}

If a function, such as the components of these objects, is a smooth function of $x$ and $y$ at the origin, this translates into a particular dependence on $X$ and $\phi$ there. We can expand such a function in Fourier modes
\begin{equation}
f\left(X\phi\right) = \sum_n f_n\brak{X,\phi} =  \sum_n \left(a_n(X) \cos n \phi + b_n(X) \sin n \phi\right),
\end{equation}
and see what happens when we impose smoothness mode by mode. The symmetries $X\rightarrow -X$ and $\phi \rightarrow \phi+\pi$ of our solutions simplify things, and mean that the scalar and tensor components consist only of even modes of the form
\begin{equation}
f_e\left(X,\phi\right) = \sum_n f_{2n}\brak{X,\phi} = \sum_n \left(A_{2n}\left(X^2\right) \cos 2 n \phi + B_{2n}\left(X^2\right) \sin 2 n \phi\right),
\end{equation}
whereas the vector components are the odd modes
\begin{equation}
f_o\brak{X,\phi} = \sum_n f_{2n+1}\brak{X,\phi}= \sum_n X \left(A_{2n+1}\left(X^2\right) \cos (2 n+1) \phi + B_{2n+1}\left(X^2\right) \sin (2 n+1) \phi \right).
\end{equation}
Note in particular that, thanks to the symmetry, the functions $A_j$ and $B_j$ depend only on $X^2$.
Since a Fourier mode like this is a smooth functions of the related Cartesian components $x$ and $y$ if it takes the form
\begin{equation}
f_n \brak{X,\phi} = X^n\left(\tilde{A}_{n}\left(X^2\right) \cos n \phi + \tilde{B}_{n}\left(X^2\right) \sin n \phi\right),
\end{equation}
all we need to check for each Fourier mode is that they fall off as $X^n$.  We can turn that on it's head and demand that if we expand our functions in powers of $X$, the $\phi$ dependence at order $X^n$ should be a combination of terms of the form $\cos m \phi$, and $\sin m \phi$ for $m \leq n$. If we show that this holds up to order $X^m$, we will have shown that the functions have well defined derivatives in $x$ and $y$ up to order $m$ at the origin. In Figure \ref{fig:originsmoothness} we check for Fourier modes that violate this condition at order $X^0$, $X$, and $X^4$. We see very fast convergence in the first two plots, and even at order $X^4$ we are still seeing fourth order convergence. This seems to indicate that the functions are well behaved at least up to the fourth derivatives at the origin.

\begin{figure}
\subfigure[Tracelessness]{\includegraphics[width=\textwidth]{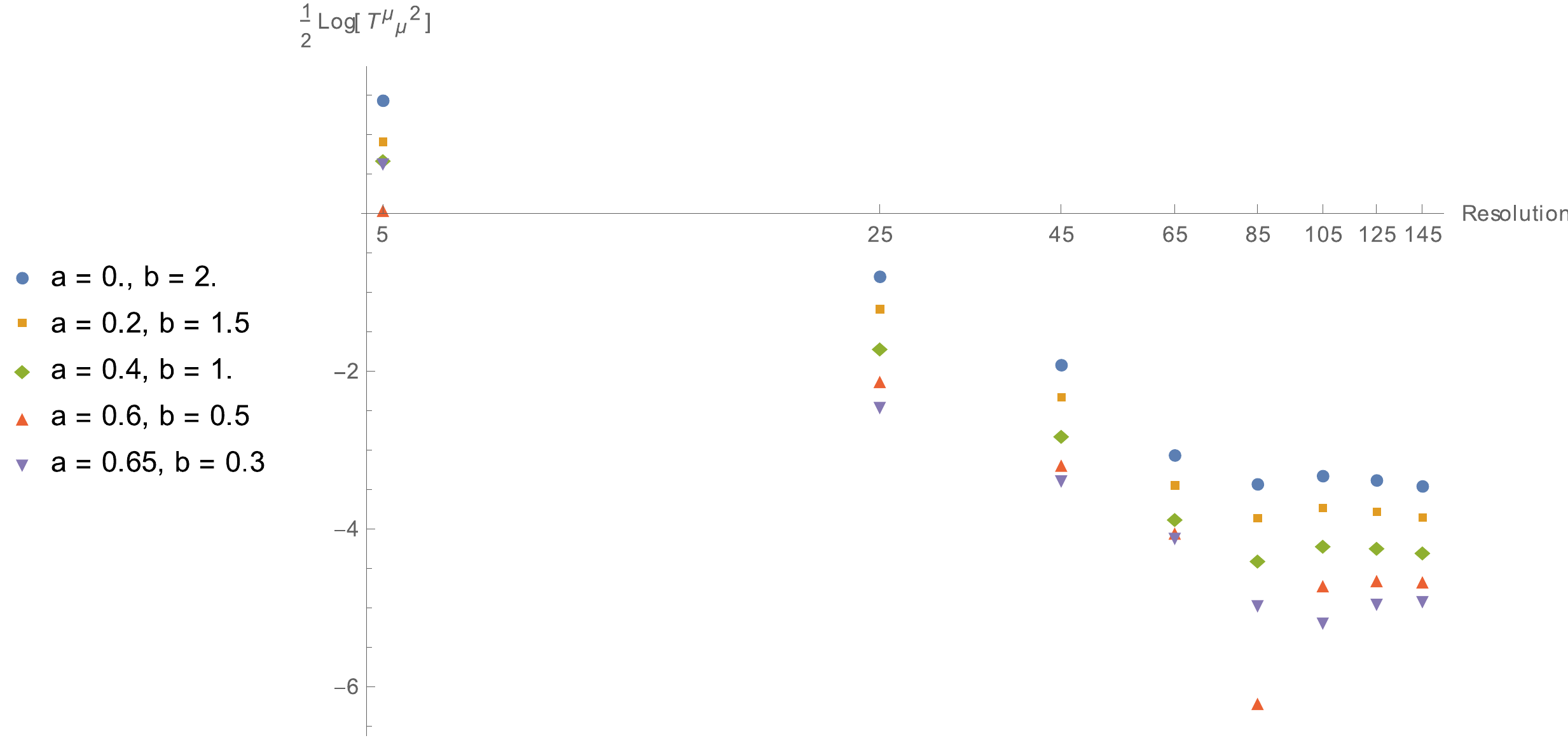}\label{sffq12}}
\subfigure[Conservation]{\includegraphics[width=\textwidth]{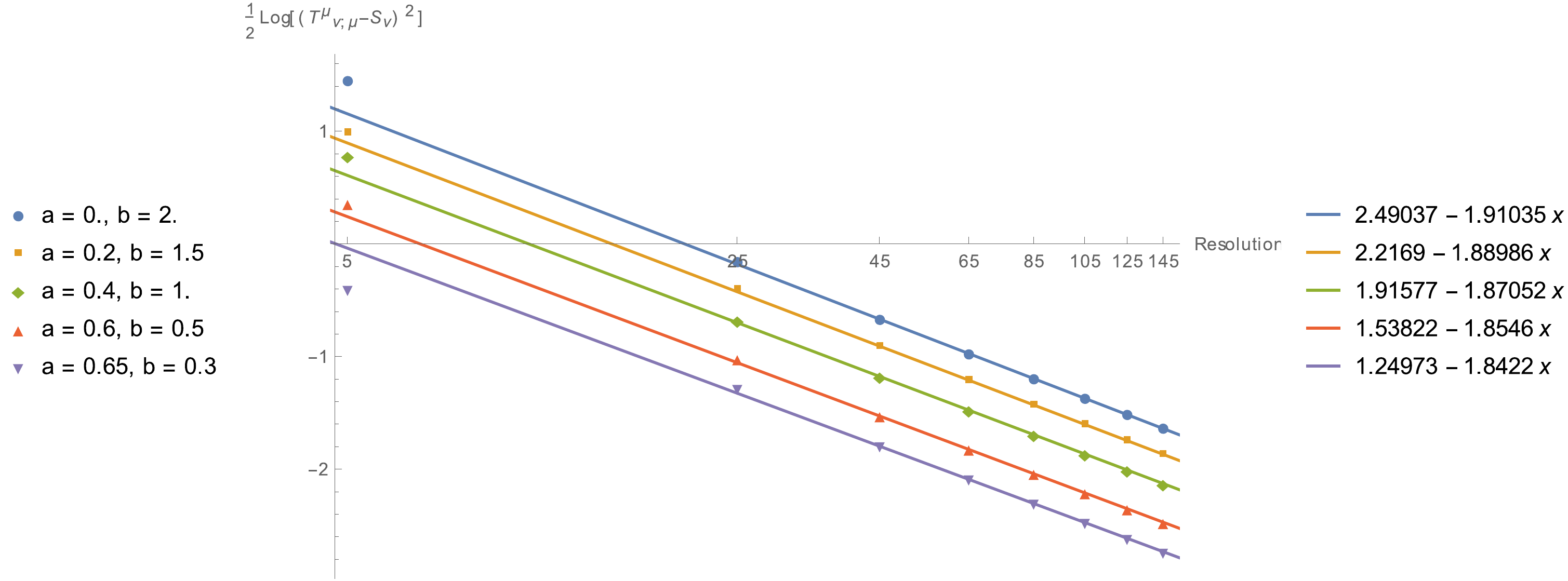}\label{sffq23}}
\caption{We plot how well the stress tensor satisfies the conservation and trace requirements. Neither converges very rapidly, but given the lack of smoothness on the boundary, we would not expect them to. }
\label{fig:stconvergence}
\end{figure}

Another check of agreement with Einstein's Equations is to check the asymptotic expansion of these solutions near the conformal boundary. To this end, we check whether the extracted stress tensor is traceless and satisfies the correct conservation law\footnote{The presence of the gauge field source term in the CFT means that the conservation law for the stress tensor becomes $\nabla_\mu T^\mu_\nu = -\frac{4}{3} F_{\mu \nu}J^\nu$, where the factor $\frac{4}{3}$ just comes from the fact that we have used a non-standard normalization of the stress tensor.}. This is shown in Figure \ref{fig:stconvergence}. We see that these quantities do not converge very rapidly, but, given the lack of smoothness on the boundary, we would not expect them to.

\section{Check of numerics against Extremal Solutions}
As discussed in Section \ref{sec:extremalcase}, when $\chi(\phi)=0$  and the gauge field is rotationally symmetric, the static bulks that preserve the symmetries of Equation \ref{eq:scalingsymmetry} are smooth 4 dimensional near-horizon geometries. These solutions are known analytically, with their form given in Equation \ref{eq:4dSolution}. We focus only on the pure gravity case here, so no gauge field, and use these solutions as a sanity check.

In this case, Weyl squared is given by
\begin{equation}
C^2 = \frac{12 \left(\psi_0^3-\psi_0\right)^2}{\psi ^6}.
\end{equation} 

In addition, the $g_{rr}$ component of the metric is invariant under the coordinate freedoms \ref{eq:coordinatefreedoms}, so we know that
\begin{equation}
g_{rr} = \psi^2.
\end{equation}

We can combine these together to get a scalar relation which should be satisfied by our solution
\begin{equation}
C^2 g_{rr}^3  = 12 \left(\psi_0^3-\psi_0\right)^2.
\end{equation}

\begin{figure}
\centering
\ssubfigure{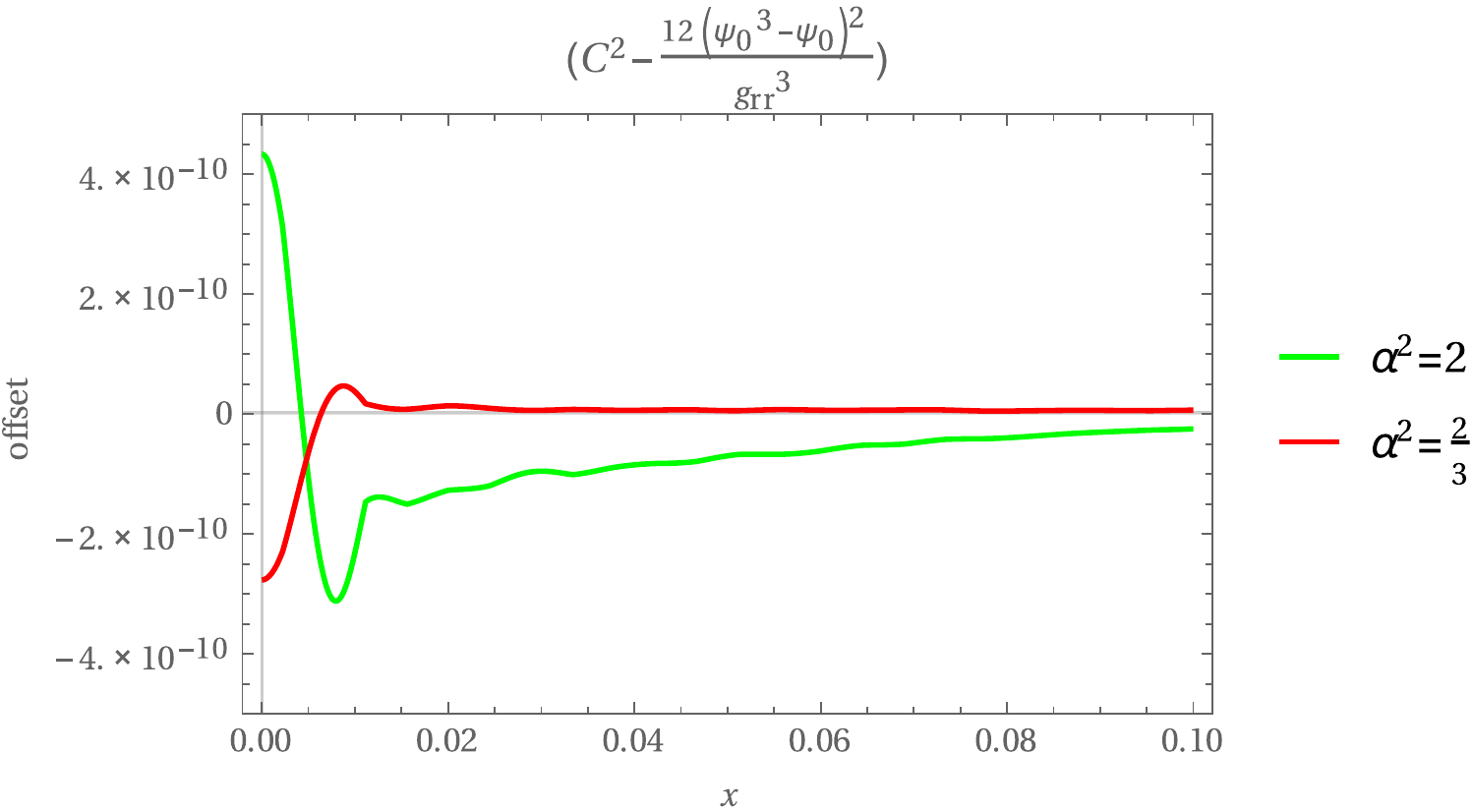}{1}{}{fig:weylerror}
\caption{Difference between the expected and numerically found value of Weyl Squared for the extremal solutions for $\alpha = \sqrt{2}$ and $\alpha = \sqrt{\frac{2}{3}}$. }
\label{fig:weylfd}
\end{figure}

The largest error we find on this comes near the axis of symmetry, and we plot a comparison for a couple of choices of $\alpha$ in Figure \ref{fig:weylfd}.

\section{Comparison with Linearised Solutions}
In \cite{Hickling2014} we presented a linearised calculation for the case of no gauge fields. We reproduce it here, with the addition of a gauge field, for completeness, and so that we can compare it to the numerical solutions we have presented. We look at linear perturbations of $AdS_4$ that preserve the static scaling symmetry. Because of the rotational symmetry of the background we're expanding about, we can consider Fourier modes and write a gauge fixed perturbation in the form

\begin{equation}
\begin{split}
 g &= \frac{\psi^2}{\rho^2}\left(-dt^2+d\rho^2\right)+ \frac{1}{\psi^2-1}d\psi^2+\left(\psi^2-1\right)d\phi^2 
 \\&+ \epsilon \cos n\phi\left( p_n(\psi)\left(\frac{-d\psi^2}{\psi^2-1}+\left(\psi^2-1\right) d\phi^2\right)\right)
 \\&+ \epsilon \cos n\phi \left(2 l_n(\psi) \frac{d \rho}{\rho} d\psi \right)
 \\&+ \epsilon \sin n \phi \left(2 f_n(\psi) \frac{d \rho}{\rho} d\phi \right)
 \\&+ \epsilon \sin n \phi \left(2 h_n(\psi) d\psi d\phi \right)\\
A &= \epsilon v_n(\psi) \cos n \phi \frac{dt}{\rho}
\end{split}
\end{equation}
where $n \ge 0$. The gauge field decouples and just satisfies Maxwell's equation on the $AdS_4$ background. The solution is 
\begin{equation}
v_n(\psi) = B_1 \left(\frac{\psi-1}{\psi+1}\right)^{n/2} + B_2 \left(\frac{\psi-1}{\psi+1}\right)^{-n/2}  ,
\end{equation}
where $B_1$ and $B_2$ are constants. This has a divergence at the origin unless $B_2=0$. This gauge field makes no linear order contribution to the stress tensor, but it leads to a charge current on the boundary
\begin{equation}
J_\mu(\phi) = -\epsilon n B_1 \cos n \phi \frac{dt}{\rho}
\end{equation}

For the gravitational perturbation, we expand Einstein's equations to linear order, the rest of the problem reduces to a single second order ode for $p_n$, with the other functions determined in terms of this, with solution
\begin{equation}
\label{eq:linsol1}
p_n(\psi) = \frac{1}{\psi(\psi^2-1)}\left( C_1 \left(\frac{\psi-1}{\psi+1}\right)^n(n\psi-1)+C_2 \left(\frac{\psi+1}{\psi-1}\right)^n(n\psi+1) \right).
\end{equation}
Requiring that the perturbation is smooth forces $C_2=0$.  The other functions are then
\begin{equation}
\label{eq:linsol2}
\begin{split}
 l_n(\psi) &= -C_1 n \psi  (\psi -1)^{\frac{n}{2}-1} (\psi +1)^{-\frac{n}{2}-1}\\
 f_n(\psi) &= C_1 \left(\frac{\psi -1}{\psi +1}\right)^{n/2} \left(n \psi +\psi ^2-1\right) \\
 h_n(\psi) &= C_1 \frac{(\psi -1)^{\frac{n}{2}-1} (\psi +1)^{-\frac{n}{2}-1} (n \psi -1)}{\psi }.
\end{split}
\end{equation}
The boundary metric gets deformed to
\begin{equation}
\label{eq:boundarylinear}
 g_\partial = \frac{-dt^2+d\rho^2}{\rho^2}+d\phi^2 +2 \epsilon C_1 \sin n \phi \frac{d\rho}{\rho}d\phi,
\end{equation}
and, through the Fefferman-Graham expansion\cite{Skenderis2000}, we can extract the boundary stress tensor
\begin{equation}
\label{eq:stresstensorlinear}
 T^{\mu}_{\phantom{\mu}\nu} = C_1 \epsilon \cos n \phi \left(
\begin{matrix}
-\frac{n^2-1}{3} & 0 & 0 \\
0 & \frac{n^2+1}{3} &0\\
0 & 0 & - \frac{2}{3}
\end{matrix} \right)
+C_1 \epsilon \sin n \phi \left(
\begin{matrix}
0 & 0 & 0 \\
0 & 0 & \frac{2n}{3}\rho \\
0 & \frac{2n}{3 \rho} & 0
\end{matrix}\right).
\end{equation}

\begin{figure}
\centering
\includegraphics[width=0.9\textwidth]{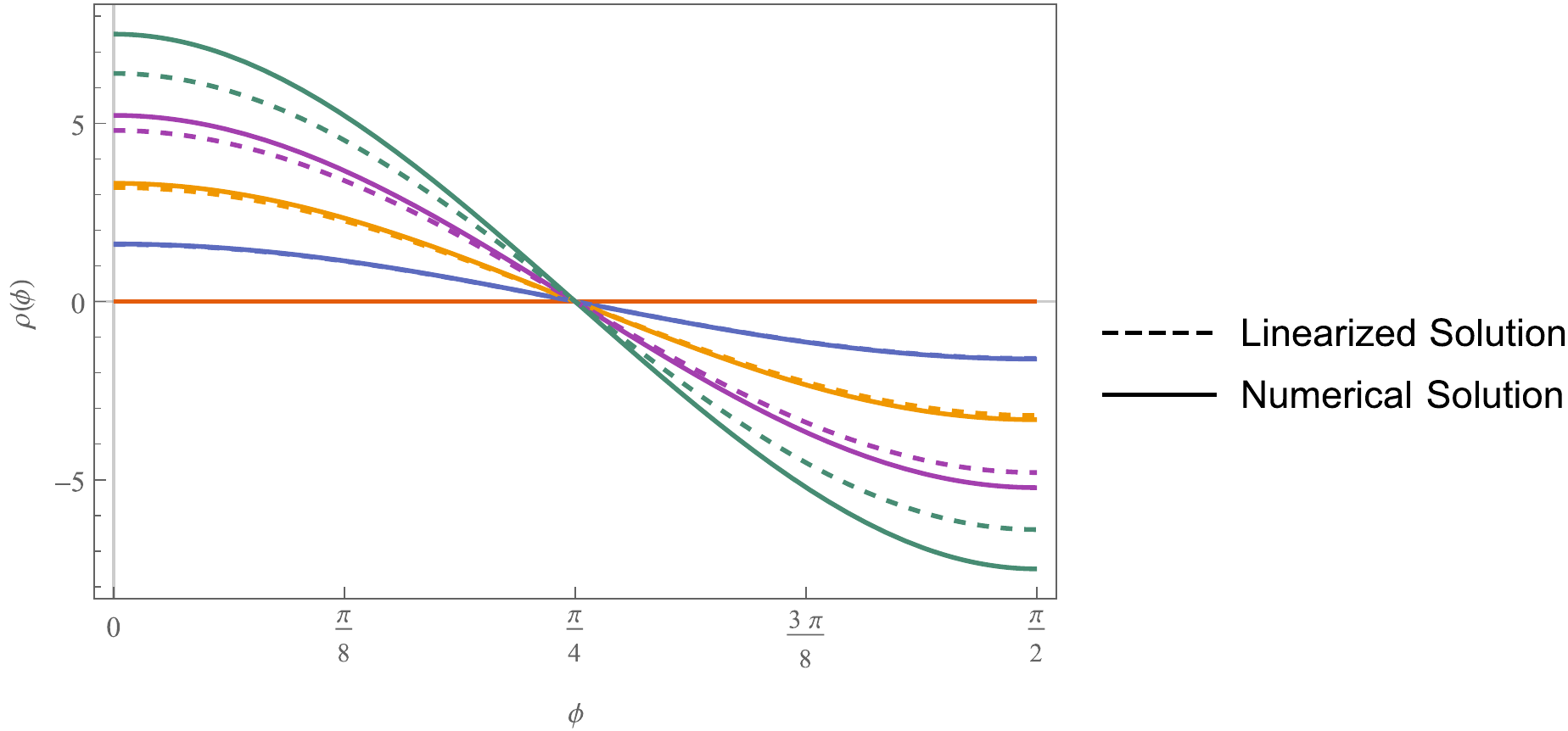}
\caption{Charge density with an electric potential source term $V(\phi) = \frac{b \cos 2\phi}{r}$ for a range of values of  $b$ compared to that of the linearized solution}
\label{fig:largescaleresponsea0lin}
\end{figure}

\begin{figure}
\centering
\includegraphics[width=0.9\textwidth]{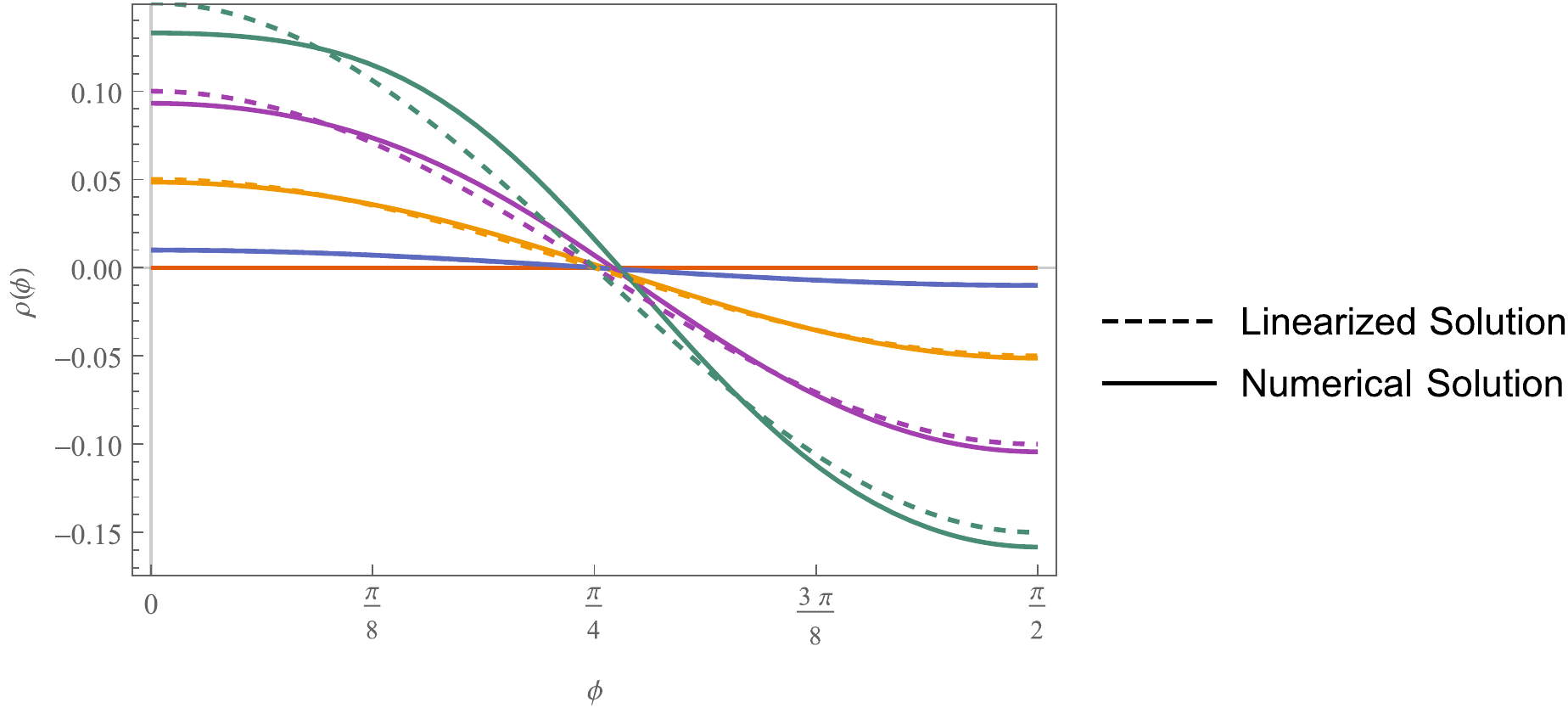}
\caption{Energy density for a boundary metric with $\chi(\phi) = \frac{\lambda \cos 2\phi}{r}$ compared with the linearized solution.}
\label{fig:energydensitylin}
\end{figure}

We compare these charge and energy densities to the ones we found numerically in Figures \ref{fig:largescaleresponsea0lin} and \ref{fig:energydensitylin}. As one would expect, initially they agree well, but there is a disagreement that grows once the perturbation grows to big.

%

\section{Logarithm in the Boundary Expansion}
\label{sec:logarithm}
\newcommand{\met}{g_{\mu \nu}}
\newcommand{\refm}{\bar{g}_{\mu \nu}}
We derive here a limitation on the smoothness of our choice of coordinates at the conformal boundary for the case of our solutions in the presence of a gauge field. We start by assuming that the metric is smooth when written in Fefferman-Graham coordinates. In these coordinates, asymptotically AdS solutions to the $3+1$ dimensional Einstein's equations can be expanded near the conformal boundary as\cite{Skenderis2000}
\begin{equation}
\met=\frac{dz^2 + \left(g_{ij}(x) + r_{ij}(x) z^2 + T_{ij}(x) z^3 +O\left(z^4\right)\right)dx^i dx^j}{z^2}.
\end{equation}
The conformal boundary is at $z=0$, $g_{ij}$ is the metric on the conformal boundary, $r_{ij}$ is determined by $g_{ij}$, but $T_{ij}$ can't be fully determined by an asymptotic expansion, and is proportional to the induced stress tensor on the boundary.

Our solutions are not written in these coordinates, but instead in coordinates fixed by the gauge condition
\begin{equation}
\label{eq:gc}
\xi^{\mu} = g^{\alpha \beta}\left(\chris{\mu}{\alpha}{\beta}-\chrisg{\mu}{\alpha}{\beta}{\bar{\Gamma}}\right).
\end{equation}
The reference connection, $\chrisg{\mu}{\alpha}{\beta}{\bar{\Gamma}}$, is derived from a reference metric $\refm$, which we fix in a advance. 

In our case, we have chosen $\met$ to itself be a solution of Einstein's equations, which means that it can also be written in Fefferman-Graham coordinates, in which it has the expansion

\begin{equation}
\refm=\frac{dz^2 + \left(\bar{g}_{ij}(x) + \bar{r}_{ij}(x) z^2 + \bar{T}_{ij}(x) z^3 +O\left(z^4\right)\right)dx^i dx^j}{z^2}.
\end{equation}
 In order to impose $\xi^\mu \to 0$ as $z \to 0$, we choose $\bar{g}_{ij} = g_{ij}$ (which implies that $\bar{r}_{ij} = r_{ij}$). If we write both $\met$ and $\refm$ in these Fefferman-Graham coordinates then we have
\begin{equation}
\met = \refm + z\left(T_{ij}(x) - \bar{T}_{ij}(x)\right) dx^i dx^j +O(z^2).
\end{equation}
Plugging this into \eqref{eq:gc} we see that
\begin{equation}
\begin{split}
\xi^z &= \frac{z^5}{2}( \bar{T} - T) + O(z^6)\\
\xi^i &= z^5\left(\nabla_j (T^{ji} - \bar{T}^{ji})-\half \nabla^i (T-\bar{T})\right)+O(z^6),
\end{split}
\end{equation}
where $T= h^{ij}T_{ij}$ and $\nabla_i$ is the covariant derivative for $h$. We see the gauge condition is satisfied at this order if $T_{ij}$ is traceless and conserved. In the absence of a bulk gauge field, this is the case. However, the addition of a bulk gauge field means that, while the CFT stress tensor is still traceless, it is no longer conserved. The boundary value of this bulk gauge field, $A_i$, is a source term in the CFT which gives rise to a charge current $J^i$. The expectation value of both this and the stress tensor are defined through the variation of the effective action
\begin{equation}
\delta W = \int \sqrt{g}\vev{ T_{ij} } \delta g^{ij} + \vev{J^i} \delta A_i.
\end{equation}
Demanding that this variation vanishes for coordinate transformations leads to the conservation law
\begin{equation}
\nabla^j T_{ij} = F_{ij}J^j.
\end{equation}
This means that in our case
\begin{equation}
\begin{split}
\xi^z &= O(z^6) \\
\xi^i &= z^5\left(F_{ij} J^j - \bar{F}_{ij}\bar{J}^j\right) + O(z^6).
\end{split}
\end{equation}
In fact, we've used the near horizon solutions as our reference metrics, and these have $\bar{F}_{ij}=0$ and $\bar{J}^i=0$. We can use this asymptotic form of $\xi^i$ in these coordinates to find an expansion for the transformation from Fefferman-Graham coordinates to the coordinates that satisfy the Harmonic gauge condition. It's found to be given by
\begin{equation}
x^i \rightarrow x^i + \frac{z^5 \log z}{5} F_{ij} J^j+O(z^6).
\end{equation}
The presence of a logarithm in this transformation inevitably translates into a logarithm term in the boundary expansion of our solutions at fifth order. This means that they cannot be more than $C^4$.

\newpage
\bibliography{./master}{}
\bibliographystyle{hunsrtnat}
\end{document}